\documentclass[aps,pra,groupedaddress,showkeys,reprint]{revtex4-2}

\usepackage{amsmath}
\usepackage{mathtools}
\usepackage{physics}
\usepackage{xfrac}
\usepackage{txfonts, comment,stmaryrd}
\usepackage[T1]{fontenc}
\usepackage{graphicx}
\usepackage{dcolumn}
\usepackage{bm}
\usepackage{hyperref}
\hypersetup{
colorlinks = true,
linkcolor  = blue,
citecolor  = blue,
urlcolor   = blue
}
\usepackage{mleftright}
\mleftright
\usepackage[usenames,dvipsnames]{color}

\begin{document}
\title{A generalized effective spin-chain formalism for strongly interacting spinor gases in optical lattice}
\author{Sagarika Basak}
\email{basak.sagarika@rice.edu}
\author{Han Pu}
\email{hpu@rice.edu}
\affiliation{Department of Physics and Astronomy, Rice University, Houston, Texas 77251, USA}

\date{\today}

\begin{abstract}
A generalized effective spin-chain model is developed for studies of strongly interacting spinor gases in a one-dimensional (1D) optical lattice. The spinor gas is mapped to a system of spinless fermions and a spin-chain. A generalized effective spin-chain Hamiltonian that acts on the mapped system is developed to study the static and dynamic properties of the spinor gas. This provides a computationally efficient alternative tool to study strongly interacting spinor gases in 1D lattice systems. This formalism permits the study of spinor gases with arbitrary spin and statistics, providing a generalized approach for 1D strongly interacting gases. By virtue of its simplicity, it provides an easier tool to study and gain deeper insights into the system. In combination with the model defined previously for continuum systems, a unified framework is developed. Studying the mapped system using this formalism recreates the physics of spinor gas in 1D lattice. Additionally, the time evolution of a quenched system is studied. The generalized effective spin-chain formalism has potential applications in the study of a multitude of interesting phenomena arising in lattice systems such as high-$T_c$ superconductivity and the spin-coherent \& spin-incoherent Luttinger liquid regimes. 
\end{abstract}


\maketitle

\section{\label{sec:level1}Introduction}
Ultracold gases are an exceptional platform to study quantum many-body physics \cite{rom_state_2004,fisher_boson_1989,jaksch_cold_1998,greiner_quantum_2002,orzel_squeezed_2001,paredes_tonksgirardeau_2004,kinoshita_observation_2004,demarco_onset_1999,truscott_observation_2001,regal_observation_2004,chin_observation_2004,zwierlein_vortices_2005}, and have potential applications in quantum optics and quantum information processing \cite{bloch_quantum_2012}. Clean (free of impurities), versatile, and extremely tunable, controllable and precise \cite{bloch_quantum_2012,gross_quantum_2017}, they have led to novel phenomena such as superfluid to Mott insulator transition \cite{bloch_quantum_2012}, Bardeen--Cooper--Schrieffer to Bose--Einstein condensate (BCS--BEC) crossover, and simulation of magnetic fields via artificial gauge fields \cite{goldman_topological_2016}. Trapping in optical lattices supplies added controllability of system parameters such as interactions, density, kinetic energy, and dimensionality \cite{bloch_many-body_2008,cazalilla_one_2011,greiner_optical_2008,senaratne_spin-charge_2022,cazalilla_ultracold_2014}. Notable are superexchange in double-well systems \cite{trotzky_time-resolved_2008}, dipole-induced spin exchange in polar lattice gases \cite{yan_observation_2013,de_paz_nonequilibrium_2013}, Harper-Hofstadter Hamiltonian \cite{aidelsburger_realization_2013,miyake_realizing_2013}, and simulation of a quantum Ising model using tilted lattices \cite{simon_quantum_2011,meinert_quantum_2013}. Quantum effects are more pronounced in reduced dimensions; one-dimensional (1D) quantum systems exhibit distinctive behaviors such as ``fermionization'' of bosons \cite{girardeau_relationship_1960} and the spin-charge separation in interacting fermions \cite{hilker_revealing_2017,auslaender_tunneling_2002,auslaender_spin-charge_2005,jompol_probing_2009,segovia_observation_1999,kim_observation_1996,kim_distinct_2006}. Experimental developments in trapping of 1D ultracold gases \cite{paredes_tonksgirardeau_2004,kinoshita_observation_2004,haller_realization_2009,moritz_confinement_2005,liao_spin-imbalance_2010}, have led to an immense interest in 1D quantum systems \cite{hilker_revealing_2017,salomon_direct_2019,basak_strongly_2021,cazalilla_one_2011,guan_fermi_2013,giamarchi_quantum_2003,paredes_tonksgirardeau_2004,kinoshita_observation_2004,haller_realization_2009,jacqmin_sub-poissonian_2011,fabbri_dynamical_2015,yang_measurement_2018,boll_spin-_2016}.   

Spinor gases, ensembles of ultracold atoms with the internal states treated as spin degrees of freedom \cite{kawaguchi_spinor_2012,stamper-kurn_spinor_2013}, are ideal to explore novel many-body physics, and immensely relevant for the study of quantum magnetism \cite{stenger_spin_1998,ciobanu_phase_2000,sadler_spontaneous_2006,huh_observation_2020,zwierlein_direct_2006,roch_quantum_2008,gemelke_situ_2009,murthy_observation_2015,williamson_coarsening_2017,williamson_universal_2016,fujimoto_unconventional_2018}, high energy physics \cite{cazalilla_ultracold_2009,gorshkov_two-orbital_2010,messio_entropy_2012,zhang_spectroscopic_2014} and quantum metrology \cite{luo_deterministic_2017,zou_beating_2018,liu_nonlinear_2022,pezze_quantum_2018,gross_nonlinear_2010}. The Hubbard model, used for the study of strongly correlated electrons in condensed matter \cite{cazalilla_ultracold_2009,gorshkov_two-orbital_2010,wu_dressed_2003,fukuhara_mott_2009,sugawa_interaction_2011}, can be realized by trapping spinor gases in optical lattices, where the correlations necessary for quantum magnetism arise from superexchange or spin-exchange interaction \cite{chomaz_dipolar_2023}. The realization of ferromagnetic interaction in bosons \cite{trotzky_time-resolved_2008,fukuhara_microscopic_2013,brown_two-dimensional_2015} and anti-ferromagnetic interaction in fermions \cite{greif_short-range_2013,hart_observation_2015,drewes_antiferromagnetic_2017,parsons_site-resolved_2016,boll_spin-_2016,cheuk_observation_2016,duan_controlling_2003,murmann_antiferromagnetic_2015,mazurenko_cold-atom_2017} have paved the way for the study of spinor gases. Spin-$\sfrac{1}{2}$ fermions with anti-ferromagnetic correlations \cite{greif_short-range_2013,hart_observation_2015,messer_exploring_2015} are linked to high-$T_c$ superconductivity \cite{norman_challenge_2011} and have the potential to depict interesting phenomena such as Stoner’s itinerant ferromagnetism \cite{drewes_antiferromagnetic_2017} and spin-incoherent Luttinger liquids \cite{dutta_non-standard_2015}. Multi-flavor Fermi--Hubbard model hosts `flavor-selective' Mott insulators \cite{vojta_orbital-selective_2010,tusi_flavour-selective_2022} linked to a new class of high-$T_c$ superconductors \cite{kamihara_iron-based_2008} and shows exotic phases \cite{hafez-torbati_competing_2019,hafez-torbati_interaction-driven_2020,nie_ferromagnetic_2017,unukovych_su4-symmetric_2021,wang_slater_2019,zhou_mott_2018,xu_interaction_2018,yamamoto_quantum_2020,keselman_emergent_2020,yao_topological_2021}. For multi-component bosons trapped in optical lattices, hyperfine states can be used to generate spin degrees of freedom for the possible exploration of novel highly entangled states \cite{yang_cooling_2020,sun_realization_2021}. 

\textit{Theoretical tools for continuum systems ---} A powerful theoretical model for 1D systems is the Bethe ansatz, uncovering interesting phenomena such as Tomonaga-Luttinger liquid (TLL) \cite{giamarchi_quantum_2003,imambekov_one-dimensional_2012,lee_thermodynamic_1988}, the Fulde-Ferrell-Larkin-Ovchinnikov (FFLO) \cite{fulde_superconductivity_1964}. Inclusion of interactions increases the Bethe ansatz complexity; circumventing this are tools such as Luttinger liquid theory \cite{voit_one-dimensional_1995}, Bosonization, Density Matrix Renormalization Group (DMRG) \cite{schollwock_density-matrix_2011,kollath_spin-charge_2005}, Cluster Perturbation Theory (CPT) \cite{senechal_spectral_2000}, and Green’s function approach \cite{economou_greens_2006}. Systems with interactions and large internal degrees of freedom are extremely challenging for theoretical investigations. In the weak interaction regime, the single-mode approximation \cite{pu_spin-mixing_1999,jie_mean-field_2020} well-describes the ground state \cite{yi_single-mode_2002,yi_quantum_2004,pu_spin-mixing_1999,ho_spinor_1998,xu_efficient_2019,chang_number_2007,evrard_many-body_2021,evrard_observation_2021,guan_tailored_2021,yi_spontaneous_2006} and excitation \cite{pu_spin-mixing_1999,ho_spinor_1998,xu_efficient_2019,santos_roton-maxon_2003}. In the strong interaction regime, systems show rich phases due to unique strongly correlated quantum effects, but investigations are computationally challenging. Under certain conditions these can be mapped to another theoretically tractable system. Notable is the Bose-Fermi Mapping \cite{girardeau_relationship_1960} of bosons in 1D with $s$-wave interaction to fermions with $p$-wave interaction \cite{girardeau_ground-state_2001,girardeau_relationship_1960,cheon_fermion-boson_1999,cheon_realizing_1998,deuretzbacher_exact_2008,guan_exact_2009,girardeau_two_2010,girardeau_tonks-girardeau_2011,fang_exact_2011}. A generalized Bose-Fermi mapping was developed for 1D continuum spinor gases \cite{yang_strongly_2015,yang_bose-fermi_2016} and subsequently to study strongly interacting pure spinor quantum gases \cite{volosniev_strongly_2014,levinsen_strong-coupling_2015,massignan_magnetism_2015,yang_bose-fermi_2016,yang_effective_2016,volosniev_simulation_2016,loft_conancruncher_2016,loft_tunable_2016,murmann_antiferromagnetic_2015}, Bose-Fermi mixtures \cite{hu_strongly_2016,hu_strongly_2016-1,bellotti_comparing_2017,deuretzbacher_spin-chain_2017}, impurity problem \cite{levinsen_strong-coupling_2015}, spinor gases with spin-dependent interactions \cite{volosniev_engineering_2015,massignan_magnetism_2015,yang_effective_2016} and spin-orbit coupling \cite{cui_spin-orbit-coupled_2014,guan_spin_2015}. 

\textit{Theoretical tools for lattice systems ---} For strongly interacting 1D spinor gases, the Hubbard model is widely used. The spin-$\sfrac{1}{2}$ Fermi--Hubbard Hamiltonian ground state wavefunction in the large on-site interaction limit is expressed as the Slater determinant of spinless fermions defining the charge degrees of freedom and the spin wavefunction, which is governed by the 1D Heisenberg model \cite{ogata_bethe-ansatz_1990}. Numerical methods for spinor gases, apart from exact diagonalization, include quantum Monte Carlo methods that are essentially exact \cite{foulkes_quantum_2001} and DMRG that captures the ground state and low-lying excited states \cite{montorsi_hubbard_1992,fazekas_lecture_1999,white_density_1992,kuhner_phases_1998,kuhner_one-dimensional_2000,rizzi_phase_2005,roy_quantum_2020}. Their main constraints are lower accuracy in the strong interaction regime (higher correlations) and being limited to smaller system sizes (computing- and memory-intensive). Approximate analytic methods devised to study the Fermi--Hubbard model at large on-site interactions include the slave particle formalism to obtain the ground state at half filling \cite{rasetti_hubbard_1991} and the t-J model \cite{edegger_gutzwillerrvb_2007}. However, the fermionic nature, strong correlations, and the need for ground state in the thermodynamic limit make the t-J model challenging \cite{putikka_limits_2000,putikka_aspects_1992,dagotto_superconductivity_1992,prelovsek_hole_1993,poilblanc_phase_1995,hellberg_phase_1997,hellberg_greens-function_2000,kohno_ground-state_1997,calandra_charge_1998}. Studies using the t-J model and the slave particle formalism have remained focused on spin-$\sfrac{1}{2}$ fermions, and for large spin fermions limited to the Heisenberg limit. There is thus a need for a generalized theoretical model for any arbitrary spinor gas applicable for bosons and fermions, and also felicitous for large spin systems at strong on-site interactions, which the present work seeks to address.  

Here, we develop a generalized effective spin-chain model to study spinor gases trapped in a one-dimensional (1D) optical lattice, realizing $N$-component Fermi--/Bose--Hubbard for fermions/bosons. This allows for the study of gases with arbitrary spin and statistics, providing a generalized approach for 1D strongly interacting gases. The spinor gas is mapped to a system of spinless fermions and a spin-chain. A generalized effective spin-chain Hamiltonian that acts on the mapped system is developed to study the spinor gas. Over existing theoretical models, this formalism is accessible (computationally and analytically easier), flexible (to system parameters), and inherently confers a vantage point to separate the effects of spin and charge degrees of freedom. Integration with the generalized Bose-Fermi mapping previously defined for a continuum system \cite{yang_bose-fermi_2016} provides a unified framework for the study of spinor gases. Using the formalism, we demonstrate the successful reproduction of the ground state of spinor gases. Furthermore, it is applied to study the time evolution of a quenched spin-$\sfrac{1}{2}$ Fermi gas in a lattice, where it captures the dynamical properties. The generalized effective spin-chain formalism can be used to study a multitude of interesting phenomena such as high-$T_c$ superconductivity, the spin-coherent \& spin-incoherent Luttinger liquid regimes.

This paper is sectioned as follows: Sec.~\ref{sec:thesystem} describes the system of spinor gas trapped in a 1D optical lattice and the mapped system of spinless fermions and a spin-chain. Sec.~\ref{sec:effspin} details the generalized effective spin-chain Hamiltonian acting on the mapped system. Sec.~\ref{sec:results} is the study of the spinor gas using the generalized effective spin-chain formalism. Sec.~\ref{sec:timeev} extends to discuss time evolution of a quenched system. Finally, Sec.~\ref{sec:summary} summarizes the results and discusses their implications and future avenues.

\section{\label{sec:thesystem}System}
We consider a 1D system of spinor gas comprising $N$-component fermions or bosons trapped in a spin-independent optical lattice. Assume low temperature and the tight-binding limit. This describes an $N$-component Fermi--Hubbard model (fermions) or $N$-component Bose--Hubbard model (bosons). 

The Fermi--Hubbard model is governed by the Hamiltonian \cite{bonnes_adiabatic_2012,messio_entropy_2012}
\begin{align}
\label{fermiH}
H_{\text{F}} = -t\sum_{i,\alpha}(c_{i,\alpha}^{\dagger}c_{i+1,\alpha}^{\vphantom{\dagger}} + \text{H.c.})
+\sum_{i,\alpha<\beta} U_{\alpha,\beta}n_{i,\alpha}n_{i,\beta},
\end{align}
where $c_{i,\alpha}$ are the fermionic annihilation operators for spin component $\alpha$, $t$ the tunneling coefficient, and $U_{\alpha,\beta}$ the on-site interaction between fermions of spin component $\alpha$ and $\beta$.

The Bose--Hubbard model is governed by the Hamiltonian 
\begin{align}
\label{boseH}
\begin{split}
H_{\text{B}} = -t\sum_{i,\alpha}(b_{i,\alpha}^{\dagger}b_{i+1,\alpha}^{\vphantom{\dagger}} + \text{H.c.})
+&\sum_{i,\alpha<\beta} U_{\alpha,\beta}n_{i,\alpha}n_{i,\beta} \\
+&\sum_{i,\alpha} \dfrac{V_{\alpha}}{2}n_{i,\alpha}(n_{i,\alpha}-1),
\end{split}
\end{align}
where $b_{i,\alpha}$ are the bosonic annihilation operators for component $\alpha$, $t$ the tunneling coefficient, $U_{\alpha,\beta}$ the on-site inter-component interaction between components $\alpha$ and $\beta$, and $V_{\alpha}$ the on-site intra-component interaction for component $\alpha$. 

In the strongly interacting limit ($U_{\alpha,\beta},\,V_{\alpha} \gg t$), these Hamiltonians (Eqns.~\ref{fermiH} \& \ref{boseH}) can be mapped to a generalized effective spin-chain Hamiltonian. 

\textit{Mapping}: The 1D spinor gas governed by the Hamiltonians defined above, can be mapped to a system of spinless fermions and a spin-chain. This mapping is hinged on a special property of 1D systems. In 1D with $M$ particles, the spatial domain can be decomposed into $M!$ spatial sectors. The wavefunction in one spatial sector carries the complete information of the total wavefunction, which can be expressed as \cite{yang_bose-fermi_2016}
\begin{equation}    
\label{twfn}
\Psi\left(\begin{aligned}x_1,x_2,\ldots,x_M,\\\sigma_1,\sigma_2,\ldots,\sigma_M\end{aligned}\right) = 
\sum_P (\pm 1)^P P\left(\Psi^1\left(\begin{aligned}x_1,x_2,\ldots,x_M,\\\sigma_1,\sigma_2,\ldots,\sigma_M\end{aligned}\right)\right),
\end{equation}
where $x_k$ and $\sigma_k$ are the position and spin of the $k^{\rm th}$ particle, $P$ represents permutation, and $\Psi^1$ is the wavefunction in spatial sector $1$ (defined by $x_1<x_2<\ldots<x_M$). The wavefunction in spatial sector $1$ can be expressed as a direct product of the spatial ($\varphi$) and spin ($\chi$) wavefunctions \cite{yang_bose-fermi_2016}:
\begin{equation}
\label{swfn}
\Psi^1\left(\begin{aligned}x_1,x_2,\ldots,x_M,\\\sigma_1,\sigma_2,\ldots,\sigma_M\end{aligned}\right) = \sum_{a,b}A_{a,b} \varphi_a(x_1,x_2,\ldots,x_M)\chi_b(\sigma_1,\sigma_2,\ldots,\sigma_M),
\end{equation}
where the superposition coefficients are given by $A_{a,b}$.
Here the spatial wavefunction is governed by the Hamiltonian defined by the tunneling of spinless fermions, and the spin wavefunction governed by a generalized effective spin-chain Hamiltonian developed in the following section.

In the following sections, we study the system in the strong on-site interaction limit, with occupancy per site $\leq 1$. For occupancy per site $>1$, the mapping can be altered to accomodate the new low energy subspace, and will be a consideration in the future.

\section{\label{sec:effspin}Generalized Effective Spin-Chain Hamiltonian}
\begin{figure}
\includegraphics{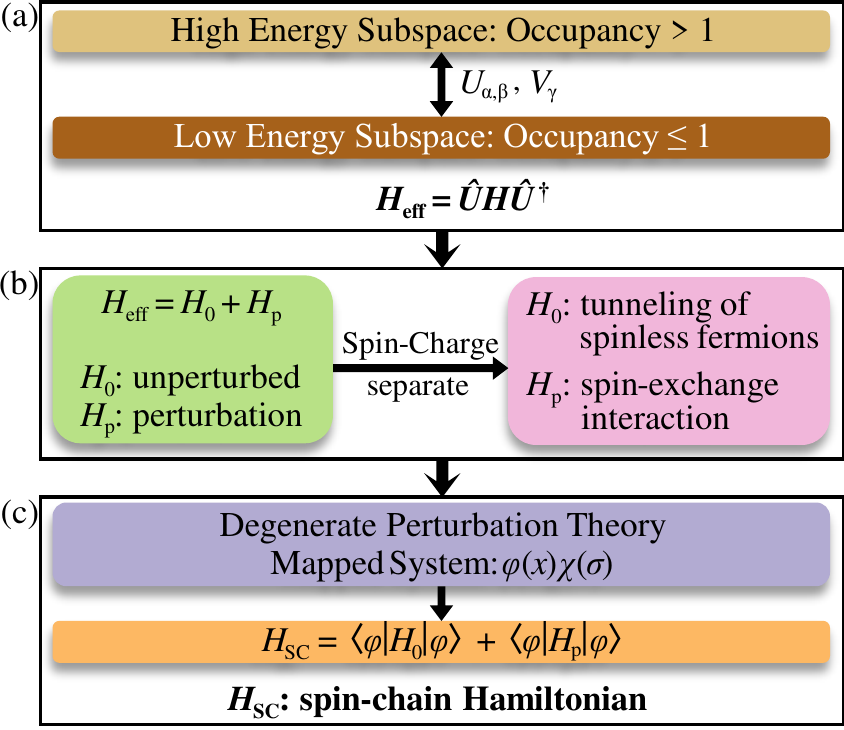}
\caption{\label{fig:subspace}(color online) Schematic of the generalized effective spin-chain Hamiltonian developed by: (a) Decoupling the low- and high-energy subspaces, roughly separated by the on-site interaction strength and coupled by the tunneling coefficients, (b) Expressing as a sum of the unperturbed Hamiltonian and a perturbation, reformulated in a simplified spin-charge separated form, and (c) Applying degenerate perturbation theory and the mapped system definition.}	
\end{figure}

Strong on-site interactions divide the Hilbert space into low- and high-energy subspaces, coupled by the tunneling term. A generalized effective spin-chain Hamiltonian is developed using a three-step protocol as in Fig.~\ref{fig:subspace}. This acts on the mapped system of spinless fermions and a spin-chain. 

The first step is to obtain an effective Hamiltonian by perturbative expansion (upto second order) at strong on-site interaction (see Appendix~\ref{sec:appendA}).
\begin{align}
H_{\text{eff}} = \mathcal{Q}^0 H\mathcal{Q}^0-\mathcal{Q}^0 H\mathcal{Q}^1\dfrac{1}{\mathcal{Q}^1 H\mathcal{Q}^1}\mathcal{Q}^1H\mathcal{Q}^0 \,, \label{eq5}
\end{align}
where $\mathcal{Q}^0$ and $\mathcal{Q}^1$ are projection operators to the low- and high-energy sectors, respectively.
The first term in Eqn.~(\ref{eq5}) represents tunneling of the spinless fermions and the second term gives rise to the spin-exchange interaction (see Appendix~\ref{sec:appendA}). 

The second step is to treat the obtained effective Hamiltonian as $H_{\text{eff}} = H_0 + H_p$, where $H_0$ ($\propto t$) is the unperturbed Hamiltonian and $H_p$ ($\propto \sfrac{t^2}{U_{\alpha,\beta}} \lor \sfrac{t^2}{V_{\alpha}}$) is treated as the perturbation. Absent perturbation (i.e., $\infty$ on-site interaction), the ground state is degenerate due to the spin degrees of freedom. The perturbation partially lifts this degeneracy. In the mapped system, the wavefunction is expressed as a direct product of spatial ($\varphi$) and spin ($\chi$) wavefunctions. The ground state of the unperturbed Hamiltonian forms the spatial wavefunction (Slater determinants) and describes the hopping of spinless fermions. The perturbation is reformulated in the simplified spin-charge separated form (see Appendices~\ref{sec:appendA} \&~\ref{sec:appendB}). 

This leads to the final step: use the degenerate perturbation theory and the mapped system wavefunction to develop the generalized effective spin-chain Hamiltonian
\begin{equation}
    H_{\text{SC}}  = \expval{H_0}{\varphi} + \expval{H_p}{\varphi} 
    = E_0 + \expval{H_p}{\varphi},
\end{equation}
where $E_0$ is the ground state energy of the unperturbed Hamiltonian.

\textit{Fermions}: Using the described protocol, the generalized effective spin-chain Hamiltonian for $N$-component fermions acting on the system of spinless fermions and a spin-chain (see Appendix~\ref{sec:appendA}) can be derived as
\begin{align}
\label{eqn:HSCF}
H_{\text{SC}}^{\text{F}}=E_0 -t^2\sum_{l=1}^{M-1}\mathcal{C}_l\left(\sum_{\alpha<\beta}^{N}\dfrac{1}{U_{\alpha,\beta}}\left(I-\mathcal{E}_{l,l+1}\right)\hat{P}_{l,l+1}^{\alpha,\beta}\right)\,,
\end{align}
where $l$ runs over the spin-chain, $M$ the number of fermions (i.e., total number of spins in the spin-chain), and $\mathcal{E}_{l,l+1}$ an exchange operator acting on spins $l$ and $l+1$ and exchanging them. An introduced projection operator $\hat{P}_{l,l+1}^{\alpha,\beta}$ restricts spins $l$ and $l+1$ to the subspace of spin components $\alpha, \beta$. 

The coupling coefficient ($\mathcal{C}_l$) for spin $l$ is given by 
\begin{align}
\label{eqn:HSCF1}
    \mathcal{C}_l = \expval{\sum_{j=1}^{L-1}\delta_{\sum_{i=1}^{j-1}n_i,l-1}\left(2\bar{n}_j^{\vphantom{\dagger}}\bar{n}_{j+1}^{\vphantom{\dagger}}-a_{j+2}^{\dagger}\bar{n}_{j+1}^{\vphantom{\dagger}}a_j^{\vphantom{\dagger}}-a_{j-1}^{\dagger}\bar{n}_j^{\vphantom{\dagger}}a_{j+1}^{\vphantom{\dagger}}\right)}{\varphi} \,,
\end{align}
where $a_{j}$ and $\bar{n}_{j}$ are the spinless fermionic annihilation and number operators respectively, $L$ is the number of lattice sites, $\delta_{\sum_{i=1}^{j-1}\bar{n}_i,l-1}$ ensures the spin $l$ corresponds to the spinless fermion on site $j$. This ensures the consistency between the tunneling Hamiltonian acting on the spinless fermions on sites $j$ and $j+1$ with the spin-chain Hamiltonian acting on spins $l$ and $l+1$. The first term $(2\bar{n}_j\bar{n}_{j+1})$ in Eqn.~(\ref{eqn:HSCF1}) reflects, within Eqn.~(\ref{eqn:HSCF}), the spin-exchange interaction between two nearest neighboring sites. This in the spin-$\sfrac{1}{2}$ space, is the familiar spin-exchange term seen in the $t-J$ Hamiltonian. The second and the third terms describe a three site hopping; with the spin-chain Hamiltonian they describe a hopping without and with spin flip in the $\alpha,\beta$ spin component subspace. In the spin-chain Hamiltonian, $\mathcal{C}_l$ includes all the charge degrees of freedom, and the operator acts on the spin wavefunction.

The generalized effective spin-chain Hamiltonian developed here assumes an extremely simple form. The action on the spin-chain as an exchange operator, provides ease of theoretical and numerical application. In comparison to the $N$-component Fermi--Hubbard acting on the original system, the formalism developed here with separation of spin and charge degrees of freedom, in addition to its simplicity, is computationally efficient.  

\textit{Bosons}: A similar protocol gives the generalized effective spin-chain Hamiltonian for $N$-component bosons acting on the system of spinless fermions and a spin-chain (see Appendix~\ref{sec:appendB}) as
\begin{widetext}
\begin{equation}
\label{eqn:HSCB}
    H_{\text{SC}}^{\text{B}} =  E_0-  t^2\sum_{l=1}^{M-1}\mathcal{C}_l\left[\sum_{\alpha<\beta}^{N}\dfrac{1}{U_{\alpha,\beta}}\left[(I+\mathcal{E}_{l,l+1})  + (\dfrac{\mu_{\alpha;\beta}+\mu_{\beta;\alpha}}{2}-1)(I + S_{l,(\alpha,\beta)}^z S_{l+1,(\alpha,\beta)}^z) + \dfrac{\mu_{\beta;\alpha}-\mu_{\alpha;\beta}}{2}(S_{l,(\alpha,\beta)}^z + S_{l+1,(\alpha,\beta)}^z) \right] \hat{P}^{\alpha,\beta}_{l,l+1}\right]\,,
\end{equation}
\end{widetext}
where aside from the terms defined for fermions, new parameters $\mu_{\alpha;\beta} =\sfrac{U_{\alpha,\beta}}{(N-1)V_{\alpha}}$ and $S_{l,(\alpha,\beta)}^z = \bar{n}_{l,\beta}-\bar{n}_{l,\alpha}$ ($\bar{n}$: number operator) are defined.
The $\mathcal{C}_l$ coefficient for both bosons and fermions are the same. Common to the fermionic (Eqn.~\ref{eqn:HSCF}) and bosonic (Eqn.~\ref{eqn:HSCB}) spin-chain Hamiltonians is the expression of identity and exchange operator through the term $(I\pm\mathcal{E}_{l,l+1})$, with minus sign for fermions representing anti-symmetric states and plus sign for bosons representing symmetric states. The bosonic spin-chain Hamiltonian has additional terms arising due to the presence of intra-component interaction. The terms are non-vanishing only when the intra-component interaction differs from the inter-component interaction or when the intra-component is spin component dependent. The Bose--Hubbard model realizes a rich phase diagram arising from the interplay of tunable inter-and intra-spin interactions \cite{duan_controlling_2003,murmann_antiferromagnetic_2015,mazurenko_cold-atom_2017,messer_exploring_2015,norman_challenge_2011,dutta_non-standard_2015,fukuhara_quantum_2013,simon_quantum_2011,meinert_quantum_2013,altman_phase_2003}.

In the limit where all the inter- and intra-component interactions are the same ($=U$), the generalized effective spin-chain Hamiltonian for fermions (Eqn.~\ref{eqn:HSCF}) and bosons (Eqn.~\ref{eqn:HSCB}) can be unified and expressed as
\begin{align}
H_{\text{SC}}^{B \lor F}=E_0 -\dfrac{t^2}{U}\sum_{l=1}^{M-1}\mathcal{C}_l\left(I\pm \mathcal{E}_{l,l+1}\right)\,.
\end{align}
The formalism developed is valid for arbitrary spin statistics as shown above, providing a superior tool to study spinor gases in the strong interaction limit. On comparison with the continuum system, it is noted that the Hamiltonians acting on the mapped systems are identical providing a universal formalism applicable to both continuum and lattice systems.  

\section{\label{sec:results}Ground State}

Previous studies on spinor gases trapped in a 1D optical lattice have extensively examined the spin and charge correlations and demonstrated interesting properties  due to multiple spin components \cite{ogata_bethe-ansatz_1990,pagano_one-dimensional_2014,manmana_su_2011,messio_entropy_2012}. Motivated by these, we explore the ground state of a spinor gas using the generalized effective spin-chain model by studying the spin and charge correlations and the associated structure factors at strong on-site interaction using DMRG for periodic boundary conditions (developed with the aid of \cite{garrison_simple-dmrgsimple-dmrg_2017, dey_efficient_2016, schollwock_density-matrix_2011}). 

\textit{Numerical Implementation}: The one-body density correlation probes the charge fluctuations and is defined in the original system as \cite{manmana_su_2011} 
\begin{align}
    C(m,n) = \dfrac{1}{N}\sum_{\sigma}C_{\sigma}(m,n) =  \dfrac{1}{N}\sum_{\sigma}\expval{c_{m,\sigma}^{\dagger}c_{n,\sigma}^{\vphantom{\dagger}}}\,,
\end{align}
where $c_{n,\sigma}$ annihilates a fermion with component $\sigma$ on site $n$.
The spin-charge separated implementation of the operator for lattice system is inspired by the definition of the momentum distribution in the mapped system for spin-$\sfrac{1}{2}$ fermions \cite{ogata_bethe-ansatz_1990}. 

\begin{figure}
\centering
\includegraphics{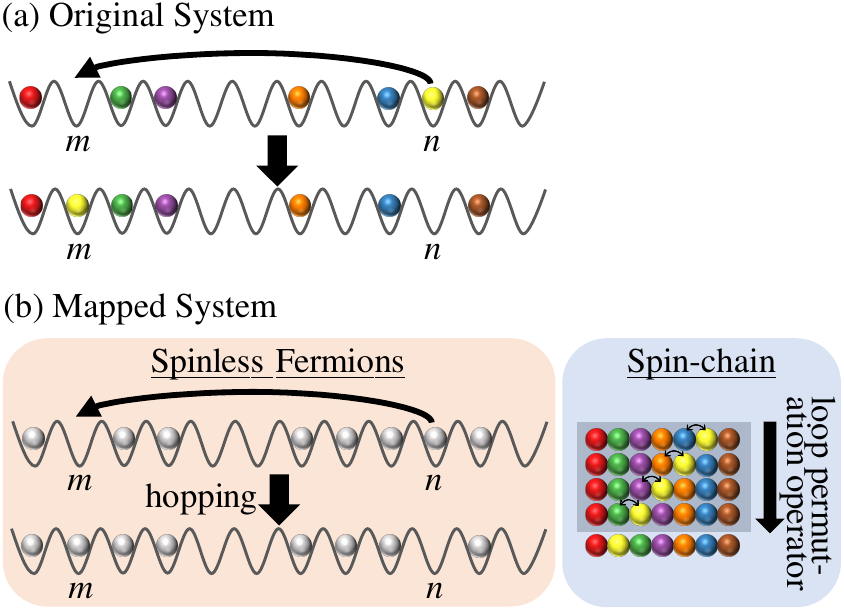}
\caption{\label{fig:schemOBDM}(color online) Schematic of the implementation of the one-body density correlation in the mapped system as hopping of spinless fermions and a loop permutation operator on the spin-chain.}	
\end{figure}

It is expressed as a hopping of spinless fermions and loop permutation operator acting on the spin-chain to represent the final spin configuration after the hopping as shown in Fig.~\ref{fig:schemOBDM}. A loop permutation operator \cite{yang_strongly_2015} acts on the spin wavefunction, which moves the $n^{\text{th}}_p$ spin to the $m^{\text{th}}_p$ via a series of exchange operators:
\begin{align}
\begin{split}
    C^{n>m}(m,n) &=  \sum_{m_p,n_p=1}^{M} (\pm 1)^{n_p-m_p}\dfrac{D^{m,n}_{m_p,n_p}}{N} \expval{\mathcal{E}_{m_p,m_p+1}...\mathcal{E}_{n_p-1,n_p}}{\chi}
    \\
    C^{n<m}(m,n) &=  \sum_{m_p,n_p=1}^{M} (\pm 1)^{m_p-n_p}\dfrac{D^{m,n}_{m_p,n_p}}{N} \expval{\mathcal{E}_{m_p,m_p-1}...\mathcal{E}_{n_p+1,n_p}}{\chi}\,,
\end{split}
\end{align}
where the plus sign is for bosons and the minus for fermions. $D^{m,n}_{m_p,n_p}$ contains the relevant charge degrees of freedom, and is expressed as 
\begin{align*}
    D^{m,n}_{m_p,n_p} = \expval{\delta^{m_p-1}_{\sum_{i_0=1}^{m-1}n_{i_0}} \delta^{n_p-1}_{\sum_{i_1=1}^{n-1}n_{i_1}} a_{m}^{\dagger}a_{n}^{\vphantom{\dagger}}}{\varphi}\,,\\
    \text{where } \delta^{m_p-1}_{\sum_{i_0=1}^{m-1}n_{i_0}} =
\begin{cases}
    1,& \text{if } \sum_{i_0=1}^{m-1}n_{i_0} = m_p-1\\
    0,& \text{otherwise}
\end{cases}\,.
\end{align*}
The $\delta$ operator ensures that the $n^{\text{th}}$ site is occupied by the $n_{p}^{\text{th}}$ spin, and after hopping the $m^{\text{th}}$ site is occupied by the $m_{p}^{\text{th}}$ spin. The Fourier component of the one-body density correlation provides an observable of great interest in the cold-atomic systems: the momentum distribution. It is defined as the occupation operator in the momentum space and expressed, assuming lattice spacing as $1$, as \cite{zhang_impenetrable_2018,ogata_bethe-ansatz_1990}
\begin{align}
    n_k =  \expval{c_{k,\sigma}^{\dagger}c_{k,\sigma}^{\vphantom{\dagger}}} = \dfrac{1}{L}\sum_{m,n=1}^{L} C(m,n) e^{ik(m-n)}\,.
\end{align}
The shape and the spread of the momentum distribution provide crucial insight into the correlations. It can be measured experimentally via time-of-flight spectra \cite{yang_one-dimensional_2011,bloch_many-body_2008}. Although long-range correlations are absent in 1D, short-range fluctuations and presence of multiple spin components provide interesting behavior, deviating from the Fermi distribution. Another observable of interest, measurable via time of flight spectra \cite{barmettler_quantum_2008}, is the structure factor associated with the spin correlation. The structure factor is the Fourier component of the two-body correlation and reveals the quantum magnetic correlations arising at strong on-site interaction. 

\noindent The diagonal two-body spin correlation is \cite{mikkelsen_relation_2023}
\begin{align}
    S(m,n) = \expval{S_m^DS_n^D}\,,
\end{align}
where $S_m^D=\sum_{\alpha,\beta} c_{m,\alpha}^{\dagger} T^D_{\alpha,\beta} c_{m,\beta}^{\vphantom{\dagger}}$, with $T^D$ being the diagonal SU($N$) generator. 

\noindent The diagonal spin correlation is expressed as \cite{mikkelsen_relation_2023,manmana_su_2011}
\begin{align}
    S(m,n) = \dfrac{1}{N(N-1)} \sum_{\alpha \neq\beta} \expval{n_{m,\alpha}n_{n,\alpha}}-\expval{n_{m,\alpha}n_{n,\beta}}\,.
\end{align}

\noindent In the mapped system, the two-body spin correlation is expressed as
\begin{align}
    &S(m,n) =  \smashoperator{\sum_{m_p,n_p=1}^{M}}\dfrac{D^{m,n}_{s,m_p,n_p}}{N(N-1)}\sum_{\alpha \neq\beta} \left(\expval{n_{m_p,\alpha}n_{n_p,\alpha}}{\chi}-\expval{n_{m_p,\alpha}n_{n_p,\beta}}{\chi}\right)\,,\\
&\text{where~} D^{m,n}_{s,m_p,n_p} = \expval{\delta^{m_p-1}_{\sum_{i_0=1}^{m-1}n_{i_0}} \delta^{n_p-1}_{\sum_{i_1=1}^{n-1}n_{i_1}} n_{m}n_{n}}{\varphi}\,.\notag
\end{align}

\noindent The structure factor associated with the two-body spin correlation is defined as
\begin{align}
    S(k) =  \dfrac{1}{L}\sum_{m,n=1}^{L} S(m,n) e^{ik(m-n)}\,.
    \label{eqn:diaSSF}
\end{align}

\begin{figure}
\centering
\includegraphics{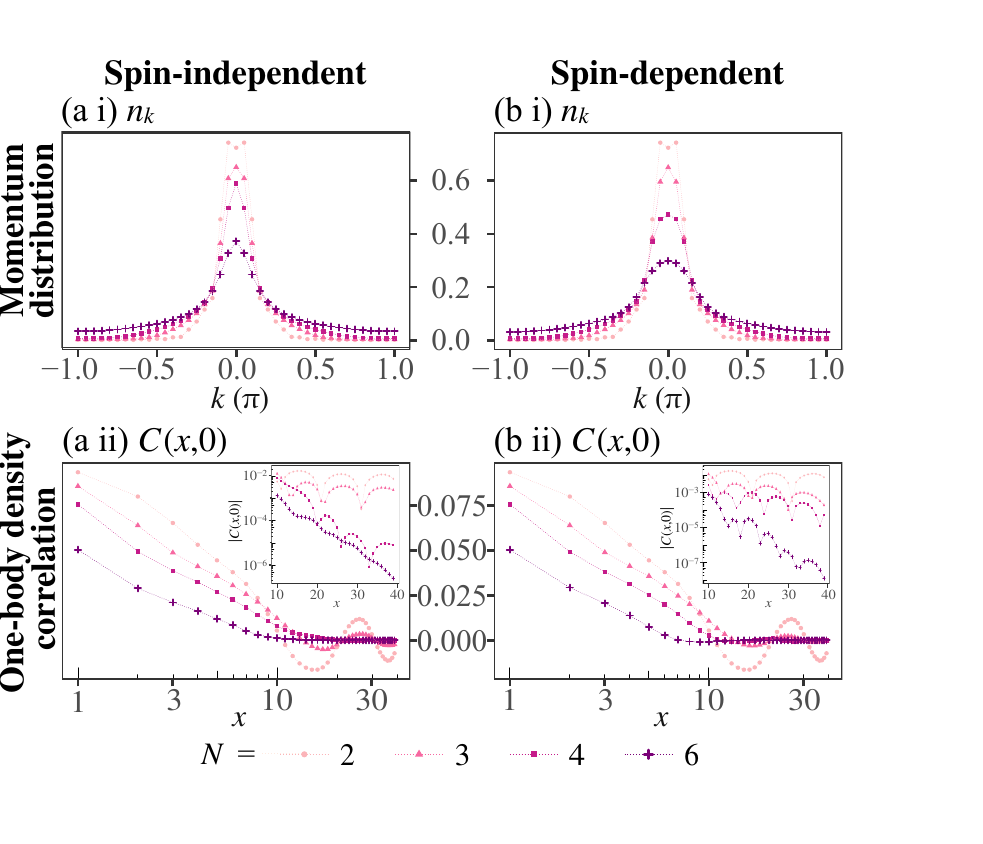}
\caption{\label{fig:fermidis}(color online) For $N$-component fermions: (i) Momentum distribution ($n_k$) as a function of momentum ($k$) and (ii) One-body density correlation ($C(x,0)$) as a function of position $x$ [\textit{Inset:} $\abs{C(x,0)}$ vs.~$x$]. Obtained for (a) spin-independent ($U$) and (b) spin-dependent ($U_{\alpha,\beta} = U(1+\sfrac{\abs{\alpha-\beta}}{(\alpha+\beta)})$) on-site interactions. Evaluated via Density Matrix Renormalization Group (DMRG) for periodic boundary conditions with system parameters: system size $L=60$, number of particles per spin component $\sfrac{M}{N}=6$, spin components $N= \{2,3,4,6\}$, and in the strong interaction region $(\sfrac{U}{t} = 20)$.}	
\end{figure}

\textit{Results}: The momentum distribution and the one-body density correlation as a function of momentum and position for $N$-component fermions obtained via the generalized effective spin-chain formalism in the presence of strong on-site spin independent and dependent interactions are shown in Fig.~\ref{fig:fermidis}. It is obtained for systems with number of spin components $N =  \{2,3,4,6\}$, size $L=60$, number of particles per component constant ($\sfrac{M}{N} = 6$) and periodic boundary conditions. Momentum distribution explores the charge fluctuations, where the correlations arising smear the known Fermi distribution. A system of spinless fermions shows a momentum distribution that increases monotonically with $k$ and a strong singularity at $k=2k_\text{F}$ ($k_\text{F}=\sfrac{\pi M}{N L}$). For spinor gases, the momentum distribution is different. The hopping of charges alters the spin configuration that results in the smearing and disappearance of the singularity at $k=2k_\text{F}$ \cite{ogata_bethe-ansatz_1990,manmana_su_2011}. For a spin-$\sfrac{1}{2}$ system, at occupancy $n<1$, a strong singularity appears at $k=k_\text{F}$ and a weak singularity at $k=3k_\text{F}$. The smearing of the distribution, as $N$ increases, results in the disappearance of the strong singularity at $k=k_\text{F}$. As the number of spin components is increased, one expects the reduction of weight at low $k$, slower decay and a monotonic broadening arising due to the fermionization of the different spin components as experimentally demonstrated \cite{pagano_one-dimensional_2014}. 

Fig.~\ref{fig:fermidis}~(a i) presents the momentum distribution with spin-independent interaction. The formalism effectively reproduces the strong singularity at $k=k_\text{F}$ for $N=2$, and its disappearance as $N$ increases. The fermionization effect is also confirmed by the observation of decreasing strength and increasing width at low values of $k$, and a slower decay as $N$ increases. It successfully captures the charge fluctuations, and the effect of spins by the consideration of occupancy $n<1$ (away from the Heisenberg limit). Studies with $N$-component fermions generally focus on spin-independent interaction with the experimental realization of trapping and cooling of alkaline earth atoms, exploring the SU($N$) symmetry of the fermionic spinor gases. However, consideration of spin-dependent interaction will create preferential spin component configurations and break the SU($N$) symmetry. The introduction of spin dependency in the on-site interaction results in further decrease of the weight at lower values of $k$ and broadening for larger $N$ as seen in Fig.~\ref{fig:fermidis}~(b i). 

Fig.~\ref{fig:fermidis}~(ii) presents the one-body density correlation, which probes the corresponding correlations in real space. It is expressed as $ \sim x^{-\sigma_\text{F}}\sin(k_\text{F}x)$, showing a power-law decay characteristic of a Luttinger liquid with an interaction-dependent exponent $\sigma_\text{F}$ \cite{kawakami_critical_1993,yamamoto_universal_2023} and $k_\text{F}$ oscillations arising due to the fermionic statistics. The power-law decay is observed over large separations, illustrated by an exponential decay in the linear-log scale in Fig.~\ref{fig:fermidis}~(ii), and logarithmic decay in log-linear scale in the insets. The $k_\text{F}$ oscillations seen in Fig.~\ref{fig:fermidis}~(ii) are better illustrated in the log-linear insets. As $N$ increases, we note that the $k_\text{F}$ oscillations become less defined and that the correlations display predominantly power-law decay. For large $N$, fermions expectedly behave as bosons \cite{pagano_one-dimensional_2014}, where the $k_\text{F}$ oscillations are absent due to Bose statistics.

\begin{figure}
\centering
\includegraphics{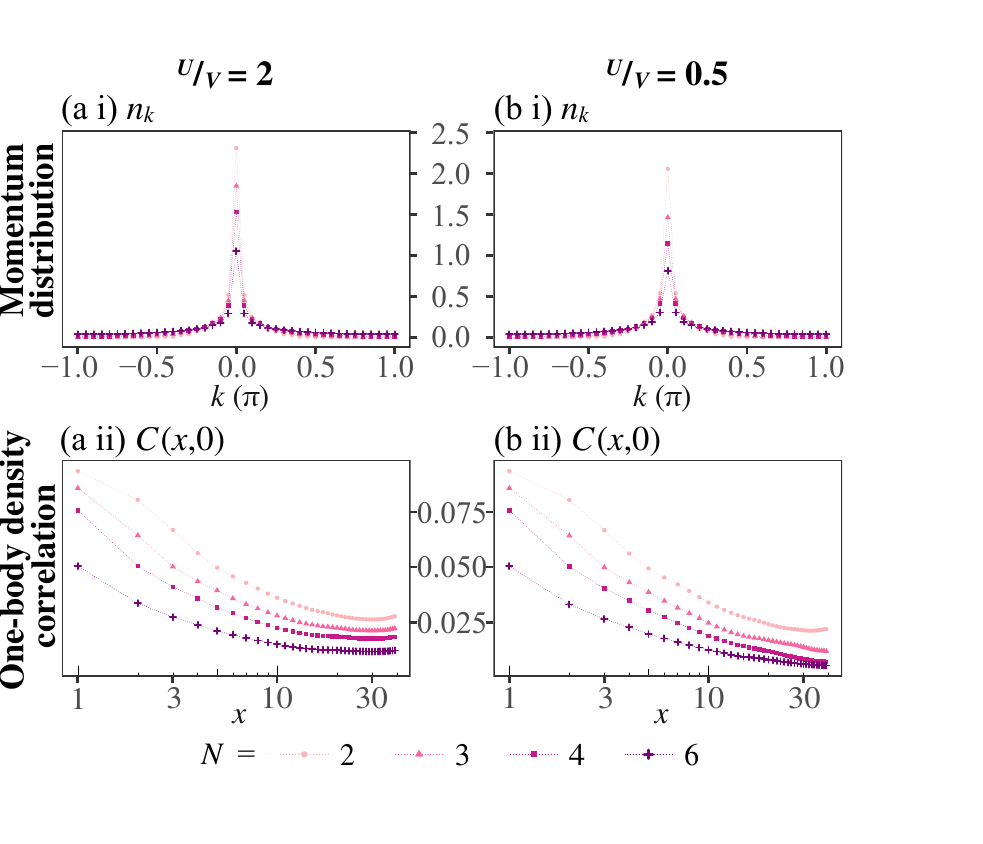}
\caption{\label{fig:bosondis}(color online) For $N$-component bosons: (i) Momentum distribution ($n_k$) as a function of momentum ($k$) and (ii) One-body density correlation ($C(x,0)$) as a function of position $x$. Obtained for inter- to intra-component on-site interaction (a) $\sfrac{U}{V}=2$ and (b) $\sfrac{U}{V}=0.5$. Evaluated via Density Matrix Renormalization Group (DMRG) for periodic boundary conditions with system parameters: size $L=60$, particles per spin component $\sfrac{M}{N}=6$, spin components $N= \{2,3,4,6\}$, and in the strong interaction region $(\sfrac{U}{t} = 20)$.}	
\end{figure}

Similar to fermions, Fig.~\ref{fig:bosondis} presents the momentum distribution and one-body density correlation obtained via the generalized effective spin-chain formalism for $N$-component bosons in the presence of strong on-site interaction. Unlike fermions, the general trend of the distribution remains insensitive to the number of components, occupancy and the strength ratio of inter- to intra-component interaction. The distribution shows a peak at $k=0$ as seen in Fig.~\ref{fig:bosondis}(i). This is expected for bosons, as higher occupancy ($>1$) for the same component is allowed. The one-body density correlation for bosons is expressed as $ \sim x^{-\sigma_\text{B}}$, where $\sigma_\text{B}$ is the interaction dependent exponent \cite{yamamoto_universal_2023}, and demonstrates approximately exponential decay in the linear-log scale as seen in Fig.~\ref{fig:bosondis}~(ii).

\begin{figure}
\centering
\includegraphics{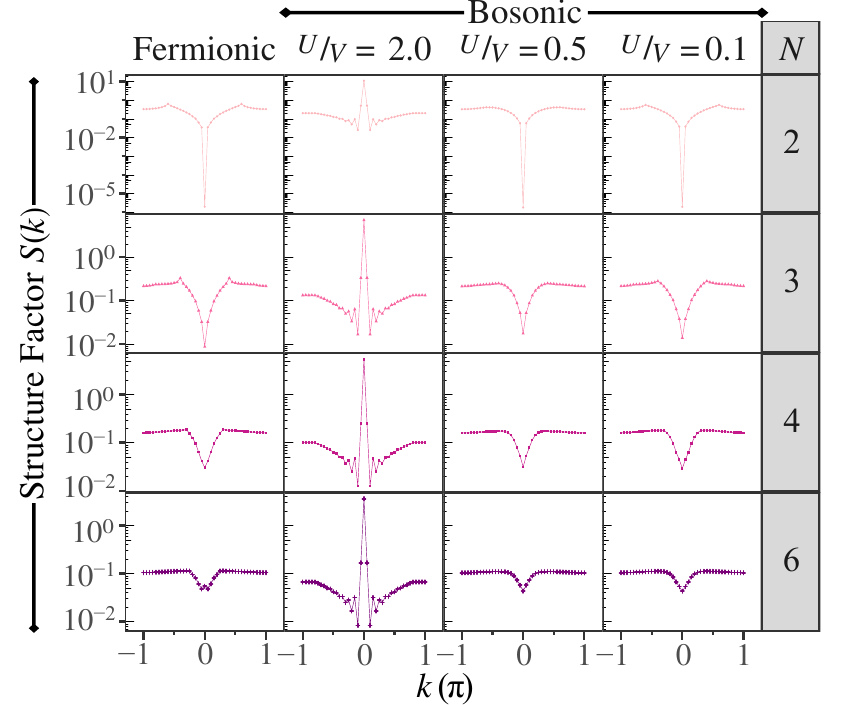}
\caption{\label{fig:fermiboscorr}(color online) Structure factor associated with diagonal two-body spin correlations ($S(k)$) as a function of momentum ($k$) for fermions with spin-independent interaction ($U$) and bosons with inter- to intra-component on-site interaction $\sfrac{U}{V}=2$, $\sfrac{U}{V}=0.5$ and $\sfrac{U}{V}=0.1$. Evaluated via Density Matrix Renormalization Group (DMRG) for periodic boundary conditions with system parameters: system size $L=60$, number of particles $M=36$,  spin components $N =  \{2,3,4,6\}$, and in the strong interaction region $(\sfrac{U}{t} = 20)$.}	
\end{figure}

Finally, let us discuss the spin Structure factor, which is an important observable that has been extensively studied and can be experimentally measured. Fig.~\ref{fig:fermiboscorr} shows the structure factor associated with the diagonal spin correlation, defined in Eqn.~(\ref{eqn:diaSSF}), for a system with components $N =  \{2,3,4,6\}$, size $L=60$ and number of particles $M=36$ obtained via the generalized effective spin-chain formalism. For fermions, a spin-independent linear behavior at small $k$ is seen in the left column of Fig.~\ref{fig:fermiboscorr}. With the occupancy considered, two peaks appear at $2k_\text{F}$ in the spin structure for all $N$, with the position of peaks being spin dependent. Increasing $N$, the peaks are shifted to a smaller $k$ and the decay of correlations at large $k$ is slower. The formalism replicates the behavior previously studied for Hubbard systems \cite{manmana_su_2011,ogata_bethe-ansatz_1990}. From the spin structure factor, the anti-ferromagnetic nature of the fermions can be confirmed by observing a dip in the correlations at $k=0$. As $N$ increases, the structure factors are non-zero at $k =0$, implying that the correlations are not purely antiferromagnetic. Larger $N$ allows a spin state to deviate from alternating spin orientation, and hence leads to non-zero contribution at $k=0$. 

The other columns in Fig.~\ref{fig:fermiboscorr} show the structure factor for bosons at different inter- to intra-component interaction strengths. Dependent on the ratio of inter- to intra-component interaction, the behavior of the spin correlations differ significantly between $\sfrac{U}{V}>1$ and $<1$. The correlations are ferromagnetic in nature for all $N$ when $\sfrac{U}{V}>1$ as seen by the strong peak at $k=0$. On the other hand, when $\sfrac{U}{V}<1$, the correlations deviate from the ferromagnetic nature displaying fermion-like behavior. This is confirmed by considering intra-component interaction much greater than inter-component interaction ($\sfrac{U}{V}=0.1$), where bosonic correlations mimic fermions. Such anti-ferromagnetic nature in spin correlations is expected, as the intra-component interaction tends to infinity. While their momentum distributions remain unalike, the bosonic spin correlations behave as fermions.

The generalized effective spin-chain Hamiltonian with the mapped system demonstrates its ability to capture the physics and reveal interesting behavior arising at occupancy $n<1$ and at strong but finite interaction strength of an arbitrary spinor gas trapped in 1D optical lattice. A comparison of the ground state correlations for different occupancy between the generalized effective spin-chain formalism and the Hubbard model for spin-$\sfrac{1}{2}$ system is provided in Appendix~\ref{sec:appendC}. In addition to being able to effectively replicate physics for arbitrary spin statistics, the separation of spin and charge degrees of freedom also provides significant computational efficiency allowing for the study of large system sizes and spin components. This method provides a superior alternative to existing theoretical tools available for spinor gases in 1D lattice system.

\section{\label{sec:timeev}Time Evolution}
In this section we discuss the time evolution of a quenched spin-$\sfrac{1}{2}$ Fermi gas trapped in a 1D optical lattice using the generalized effective spin-chain Hamiltonian. The dynamics are compared to that obtained from the original Fermi--Hubbard Hamiltonian. The initial state and the dynamics are obtained using time-dependent variational principle method (developed with input from \cite{mendl_pytenet_2018}) for a system size of $L=14$ and number of particles $M=8$ in the presence of an additional harmonic trap, and with strong on-site interaction.

\textit{Initial state}: Consider an extension to the Fermi--Hubbard Hamiltonian, with the introduction of a spin-dependent magnetic gradient. The extended Fermi--Hubbard Hamiltonian is defined as 
\begin{align}
    H_0 =-t\sum_{\expval{i,j},\sigma}(c_{i,\sigma}^{\dagger}c_{j,\sigma}^{\vphantom{\dagger}}+\text{H.c.})+\sum_i \left( U n_{i,\uparrow}n_{i,\downarrow} + V_h x_i^2 n_i + h_z x_i\sigma_i^z\right) \,,
\end{align}
where $x_i = \sfrac{(L+1)}{2}-i$ denotes the position of site $i$ from the center of the chain, local density $n_i =\sum_{\sigma}n_{i,\sigma}$, and $\sigma_i^z = n_{i,\uparrow}-n_{i,\downarrow}$ defines the local spin order along $z$. Aside from the tunneling $(t)$ and on-site interaction $(U)$, the Hamiltonian includes an additional harmonic trap $(V_h)$ and a spin-dependent magnetic gradient $(h_z)$. The ground state of this extended Fermi--Hubbard Hamiltonian at strong on-site interaction is chosen as the intial state for evolution.

In the generalized effective spin-chain formalism, appropriate Hamiltonians are defined to produce equivalent charge and spin initial states. 
The charge initial state is taken as the ground state of the Hamiltonian 
\begin{align}
    H^{\text{C}}_0 = -t\sum_{\expval{i,j}}(a_{i}^{\dagger}a_{j}^{\vphantom{\dagger}}+\text{H.c.}) +\sum_i V_h x_i^2 \bar{n}_i -h_z\sum_i \abs{x_i}\bar{n}_i
    \label{eqn:tc0}
\end{align}
and the spin initial state is the ground state of the Hamiltonian 
\begin{align}
    H^{\text{S}}_0 =-\dfrac{t^2}{U}\sum_l\mathcal{C}_l\left(I-\mathcal{E}_{l,l+1}\right) + h_z \sum_l \mathcal{D}_l\sigma_l^z\,.
    \label{eqn:tsc0}
\end{align}
The coefficients $\mathcal{C}_l$ are given in Eqn.~(\ref{eqn:HSCF1}) and 
\begin{align}
    \mathcal{D}_l = \expval{\sum_i \delta_{\sum_{j=1}^{i-1}n_j,l-1} x_i \bar{n}_i}{\varphi}\,,
\end{align}
where $\varphi$ is the initial charge ground state. Eqns.~(\ref{eqn:tc0}) \& (\ref{eqn:tsc0}) thus create an initial state in the mapped system, which is equivalent to the original system. 

\textit{Evolution}: To obtain the quench, prior to the evolution, the previously introduced spin-dependent magnetic gradient is removed while the harmonic trap is retained. 

\begin{figure}
\centering
\includegraphics{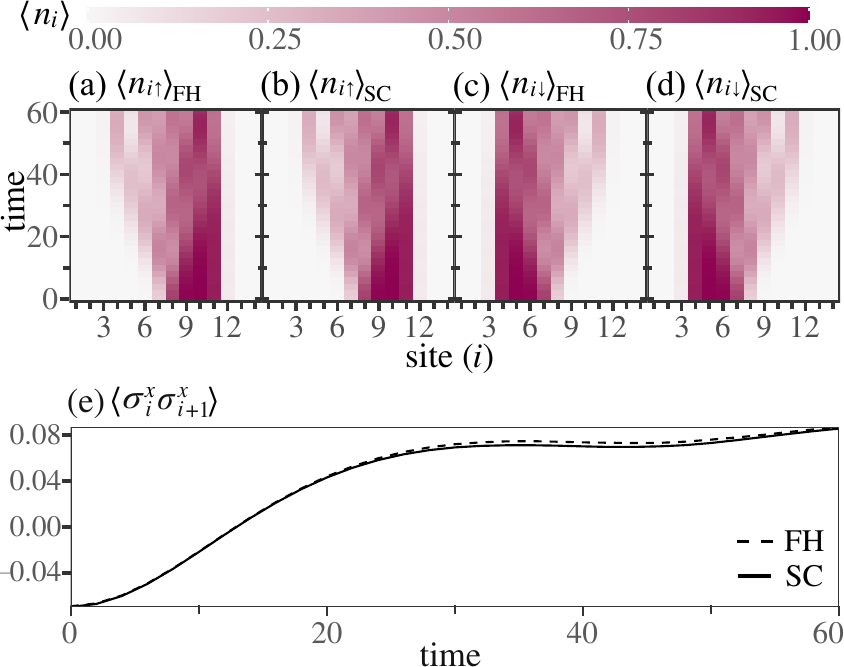}
\caption{\label{fig:spinTE}(color online) (a--d) Time evolution of spin population obtained via (a $\&$ c) Fermi--Hubbard Hamiltonian and (b $\&$ d) Spin-Chain for sites 1--14. (e) Time evolution of the nearest neighbor spin correlation along $x$ obtained via Fermi--Hubbard and via spin-chain Hamiltonian. Evaluated for spin-$\sfrac{1}{2}$ fermions via time-dependent variational principle method for open boundary conditions in the presence of a harmonic trap ($V_h = 0.5t$) with spin-dependent magnetic gradient ($h_z = 0.1t$) and system parameters: system size $L=14$, number of particles $M=8$, and in the strong interaction region $(\sfrac{U}{t} = 40)$. Time is in units of $\sfrac{1}{t}$.}
\end{figure}

\textit{Results}: The evolution of the spin populations and correlation obtained from both models are compared in Fig.~\ref{fig:spinTE}. The generalized effective spin-chain formalism is able to replicate the population and correlation dynamics obtained from Fermi--Hubbard Hamiltonian of the quenched system in the presence of a harmonic trap. Additionally, this formalism computationally outperforms the Fermi--Hubbard Hamiltonian, being $\approx 10$ times faster, and is capable of accommodating much larger system sizes with a smaller memory footprint. 

\begin{figure}
\centering
\includegraphics{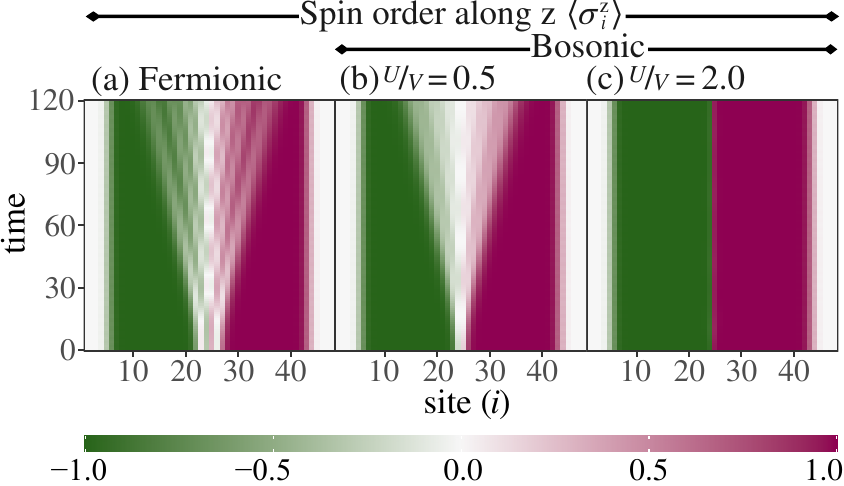}
\caption{\label{fig:ordTE_L}(color online) Time evolution of the spin order along $z$ for (a) fermions, (b) bosons with inter- to intra-component on-site interaction $\sfrac{U}{V}=0.5$ and (c) bosons with $\sfrac{U}{V}=2$ obtained via the generalized effective spin-chain formalism for sites 1--48. The populations are evaluated for spin-$\sfrac{1}{2}$ spinor gas via time-dependent variational principle method for open boundary conditions with system parameters: $V_h = 0.04t$, $h_z = 0.03t$, $L=48$, $M=38$, and in the strong interaction region $(\sfrac{U}{t} = 40)$. Time is in units of $\sfrac{1}{t}$.}
\end{figure}

As a demonstration of its computing efficiency, the dynamics of the quenched system is obtained for larger system size $L=48$ and number of particles $M=38$ for spin-$\sfrac{1}{2}$ fermions and bosons at on-site inter- to intra-component interaction strengths $\sfrac{U}{V}=2$ and $0.5$. The evolution of the spin order along $z$ is shown in Fig.~\ref{fig:ordTE_L}. The spin order for bosonic system at $\sfrac{U}{V}=2$ remains nearly stationary owing to the initial state considered and the preference for ferromagnetic ordering. The evolution of the spin order for fermions and bosons at $\sfrac{U}{V}=0.5$ show a spread in the region with $0$ order. This is expected as both systems prefer anti-ferromagnetic ordering.    

\section{\label{sec:summary}Summary}
In conclusion, we have developed a generalized effective spin-chain Hamiltonian acting on a system of spinless fermions and spin-chain to represent the physics of a spinor gas trapped in a one-dimensional optical lattice at strong on-site interaction. The developed model is valid for arbitrary spin and statistics, providing a generalized approach, and has an identical formalism as the continuum system providing a universal approach to study spinor gases in 1D systems. The separation of spin and charge degrees of freedom, and the simple nature of the Hamiltonian make it computationally efficient, allowing for the study of larger system sizes and longer time evolution using this formalism. We employ Density Matrix Renormalization Group (DMRG) to study the ground state of $N$-component fermions and $N$-component bosons. With the existing theoretical and numerical tools, studying the physics of arbitrary spinor gases for components greater than two is challenging. Here, the developed formalism effectively reproduces the physics of the spinor gas at strong on-site interaction. Using the time-dependent variational principle, the time evolution of a quenched spin-$\sfrac{1}{2}$ Fermi gas in a lattice is studied. The mapped system captures the dynamical properties of the original system, producing results consistent with the Fermi--Hubbard model. Additionally, as an illustration of the efficiency of the generalized effective spin-chain formalism, the dynamics of the spin-$\sfrac{1}{2}$ quenched fermionic and bosonic systems are presented for a large system size, which is conventionally intractable. The generalized effective spin-chain model provides an efficient tool to gain new insights into the equilibrium and non-equilibrium physics of spinor gases. 
 
\textit{Future Avenues:} Our work provides a unique, generalized, and universal approach to studying spinor gases in a one-dimensional system. Potential applications of the formalism are to provide new insights into the multi-component spinor gas such as the development of exotic phases at $N>2$, the correlation dependence on occupancy, effect of doping and consideration of finite temperature. It can be extended to consider two tilted 1D optical lattices parallel to each other to obtain the Su--Schrieffer--Heeger (SSH) model, development of a symmetry protected topological phase, and the effect of the number of spin components on this phase. Additionally, the consideration of disorder in the formalism to study localization and quantum scarring will be of interest. The generalized effective spin-chain formalism is useful in the study of a multitude of interesting phenomena arising in lattice systems such as high-$T_c$ superconductivity and the spin-coherent \& spin-incoherent Luttinger liquid regimes. 

\begin{acknowledgments}
We thank Prof.~Xiwen Guan for many useful discussions. This study was supported by the National Science Foundation under grant PHY-2207283 and the Welch Foundation under Grant No.~C-1669. This work was supported in part by the Big-Data Private-Cloud Research Cyberinfrastructure MRI-award funded by NSF under grant CNS-1338099 and by Rice University's Center for Research Computing (CRC).
\end{acknowledgments}

\appendix
\section{\label{sec:appendA}Effective Spin-Chain Hamiltonian: Fermions}
Consider a system of $N$-component fermions in a 1D optical lattice. Let the number of fermions be $M$ and the number of lattice sites $L$. Define complementary subspaces: $H_{\mathcal{Q}^0}$ (low-energy: occupancy $\leq$ 1) with projector $\mathcal{Q}^0$ and $H_{\mathcal{Q}^1}$ (high-energy: occupancy $>$ 1) with $\mathcal{Q}^1(=1-\mathcal{Q}^0)$. The effective Hamiltonian (up to second-order) can be expressed as \cite{cazalilla_one-dimensional_2003}
\begin{align}
H_{\text{eff}} = \mathcal{Q}^0 H\mathcal{Q}^0-\mathcal{Q}^0 H\mathcal{Q}^1\dfrac{1}{\mathcal{Q}^1 H\mathcal{Q}^1}\mathcal{Q}^1H\mathcal{Q}^0\,. \label{eqn:A1}
\end{align}
The hopping of the fermions contributes the zeroth-order.
\begin{align*}
\mathcal{Q}^0 H\mathcal{Q}^0 = -t\sum_{i,\alpha}\mathcal{Q}^0(c_{i,\alpha}^{\dagger}c_{i+1,\alpha}^{\vphantom{\dagger}}+c_{i+1,\alpha}^{\dagger}c_{i,\alpha}^{\vphantom{\dagger}})\mathcal{Q}^0
\end{align*}
Define projected fermions to restrict occupancy $\leq$ 1: annihilation operator $a_{i,\alpha} = \prod_{\beta \neq \alpha}c_{i,\alpha}(1-n_{i,\beta})$, creation operator $a_{i,\alpha}^{\dagger} = \prod_{\beta \neq \alpha}c_{i,\alpha}^{\dagger}(1-n_{i,\beta})$ and number operator $\bar{n}_{i,\alpha} = a_{i,\alpha}^{\dagger}a_{i,\alpha}^{\vphantom{\dagger}} = \prod_{\beta \neq \alpha}c_{i,\alpha}^{\dagger}c_{i,\alpha}^{\vphantom{\dagger}}(1-n_{i,\beta}^{\vphantom{\dagger}}) = \prod_{\beta \neq \alpha}n_{i,\alpha}(1-n_{i,\beta})$.
\begin{align*}
\mathcal{Q}^0 H\mathcal{Q}^0 = -t\sum_{i,\alpha}(a_{i,\alpha}^{\dagger}a_{i+1,\alpha}^{\vphantom{\dagger}}+a_{i+1,\alpha}^{\dagger}a_{i,\alpha}^{\vphantom{\dagger}})
\end{align*}
As the hopping preserves the spin components over which the expression is summed over, the spin symbols are removed. 
\begin{align}
\mathcal{Q}^0 H\mathcal{Q}^0 = -t\sum_{i}(a_{i}^{\dagger}a_{i+1}^{\vphantom{\dagger}}+a_{i+1}^{\dagger}a_{i}^{\vphantom{\dagger}})
\label{eqn:A2}
\end{align}
The second-order term $\mathcal{Q}^0 H\mathcal{Q}^1\dfrac{1}{\mathcal{Q}^1 H\mathcal{Q}^1}\mathcal{Q}^1H\mathcal{Q}^0$ is
\begin{flalign*}
= &\sum_{\expval{i,j,k},\alpha,\beta}t^2\mathcal{Q}^0(c_{i,\alpha}^{\dagger}c_{j,\alpha}^{\vphantom{\dagger}})\dfrac{\mathcal{Q}^1}{\mathcal{Q}^1 H\mathcal{Q}^1}(c_{j,\beta}^{\dagger}c_{k,\beta}^{\vphantom{\dagger}})\mathcal{Q}^0  &&
\end{flalign*}
\begin{flalign*}
= &\sum_{\expval{i,j,k},\alpha=\beta}t^2\mathcal{Q}^0(c_{i,\alpha}^{\dagger}c_{j,\alpha}^{\vphantom{\dagger}})\dfrac{\sum_{\gamma \neq \alpha} n_{j,\gamma}}{U_{\alpha,\gamma}}(c_{j,\alpha}^{\dagger}c_{k,\alpha}^{\vphantom{\dagger}})\mathcal{Q}^0 + &&
\\ &\sum_{\expval{i,j,k},\alpha \neq \beta} t^2\mathcal{Q}^0 (c_{i,\alpha}^{\dagger}c_{j,\alpha}^{\vphantom{\dagger}})\dfrac{1}{U_{\alpha,\beta}}(c_{j,\beta}^{\dagger}c_{k,\beta}^{\vphantom{\dagger}})\mathcal{Q}^0 &&
\\ = &\sum_{\expval{i,j,k},\alpha \neq \gamma}t^2\mathcal{Q}^0(c_{i,\alpha}^{\dagger}c_{j,\alpha}^{\vphantom{\dagger}})\dfrac{n_{j,\gamma}}{U_{\alpha,\gamma}}(c_{j,\alpha}^{\dagger}c_{k,\alpha}^{\vphantom{\dagger}})\mathcal{Q}^0 + &&
\\ &\sum_{\expval{i,j,k},\alpha \neq \beta} t^2\mathcal{Q}^0 (c_{i,\alpha}^{\dagger}c_{j,\alpha}^{\vphantom{\dagger}})\dfrac{1}{U_{\alpha,\beta}}(c_{j,\beta}^{\dagger}c_{k,\beta}^{\vphantom{\dagger}})\mathcal{Q}^0 &&
\\ = &\sum_{\expval{i,j,k},\alpha \neq \gamma}t^2\mathcal{Q}^0(c_{i,\alpha}^{\dagger}(1-n_{j,\alpha}^{\vphantom{\dagger}})\dfrac{n_{j,\gamma}}{U_{\alpha,\gamma}}c_{k,\alpha}^{\vphantom{\dagger}})\mathcal{Q}^0 +  &&
\\ & \sum_{\expval{i,j,k},\alpha \neq \beta} t^2\mathcal{Q}^0 (c_{i,\alpha}^{\dagger}c_{j,\alpha}^{\vphantom{\dagger}})\dfrac{1}{U_{\alpha,\beta}}(c_{j,\beta}^{\dagger}c_{k,\beta}^{\vphantom{\dagger}})\mathcal{Q}^0 &&
\\ \intertext{Projected fermion annhilation ($a_{i,\alpha}$) and number operators $\bar{n}_{i,\alpha}$ are introduced.}
= &\sum_{\expval{i,j,k},\alpha \neq \gamma}\dfrac{t^2}{U_{\alpha,\gamma}}(a_{i,\alpha}^{\dagger}\bar{n}_{j,\gamma}^{\vphantom{\dagger}}a_{k,\alpha}^{\vphantom{\dagger}}) - \sum_{\expval{i,j,k},\alpha \neq \beta} \dfrac{t^2}{U_{\alpha,\beta}} (a_{i,\alpha}^{\dagger}a_{j,\beta}^{\dagger}a_{j,\alpha}^{\vphantom{\dagger}}a_{k,\beta}^{\vphantom{\dagger}}) &&
\\ = &\underbrace{\sum_{\expval{i,j},\alpha \neq \beta} \left( \dfrac{t^2}{U_{\alpha,\beta}}(\bar{n}_{i,\alpha}^{\vphantom{\dagger}}\bar{n}_{j,\beta}^{\vphantom{\dagger}} - a_{i,\alpha}^{\dagger}a_{i,\beta}^{\vphantom{\dagger}}a_{j,\beta}^{\dagger}a_{j,\alpha}^{\vphantom{\dagger}}) \right)}_{\text{\textcircled{\scriptsize{1}}}} &&
\\ &\phantom{\qquad\qquad}\underbrace{-\sum_{\expval{i,j,k \neq i},\alpha \neq \beta} \left( \dfrac{t^2}{U_{\alpha,\beta}}(a_{i,\alpha}^{\dagger}\bar{n}_{j,\beta}^{\vphantom{\dagger}}a_{k,\alpha}^{\vphantom{\dagger}}-a_{i,\alpha}^{\dagger}a_{j,\beta}^{\dagger}a_{j,\alpha}^{\vphantom{\dagger}}a_{k,\beta}^{\vphantom{\dagger}}) \right)}_{\text{\textcircled{\scriptsize{2}}}} &&
\end{flalign*}

There are two second-order or virtual hopping processes that contribute to and form the effective spin-chain Hamiltonian. Process \text{\textcircled{\scriptsize{1}}} involves only two nearest neighboring sites similar to the spin-exchange process seen in the $t-J$ Hamiltonian. Process \text{\textcircled{\scriptsize{2}}} describes hopping between three nearest neighboring sites without and with spin flip.

To simplify the expressions further, we define spin operators in the $\alpha, \beta$ subspace as
\begin{align}
\begin{split}
   & n_{i,(\alpha,\beta)} = \bar{n}_{i,\beta}+\bar{n}_{i,\alpha}\,, 
\\ & S_{i,(\alpha,\beta)}^z = \bar{n}_{i,\beta}-\bar{n}_{i,\alpha}\,,
\\ & S_{i,(\alpha,\beta)}^x = a_{i,\beta}^{\dagger}a_{i,\alpha}^{\vphantom{\dagger}}+a_{i,\alpha}^{\dagger}a_{i,\beta}^{\vphantom{\dagger}}\,, 
\\ & S_{i,(\alpha,\beta)}^y = -i \left(a_{i,\beta}^{\dagger}a_{i,\alpha}^{\vphantom{\dagger}}-a_{i,\alpha}^{\dagger}a_{i,\beta}^{\vphantom{\dagger}}\right)\,.
\label{eqn:A4}
\end{split}
\end{align}

These processes are derived and expressed as follows:
\begin{flalign*}
\text{\textcircled{\scriptsize{1}}}: &\sum_{\expval{i,j},\alpha \neq \beta} \left( \dfrac{t^2}{U_{\alpha,\beta}}\left(\bar{n}_{i,\alpha}^{\vphantom{\dagger}}\bar{n}_{j,\beta}^{\vphantom{\dagger}} - a_{i,\alpha}^{\dagger}a_{i,\beta}^{\vphantom{\dagger}}a_{j,\beta}^{\dagger}a_{j,\alpha}^{\vphantom{\dagger}}\right) \right) &\\ 
&=\sum_{i,\alpha < \beta} \dfrac{2t^2}{U_{\alpha,\beta}}\left(\bar{n}_{i,\alpha}^{\vphantom{\dagger}}\bar{n}_{i+1,\beta}^{\vphantom{\dagger}} + \bar{n}_{i,\beta}^{\vphantom{\dagger}}\bar{n}_{i+1,\alpha}^{\vphantom{\dagger}}\right.&\\
&\phantom{\qquad\qquad\quad} \left. -a_{i,\alpha}^{\dagger}a_{i,\beta}^{\vphantom{\dagger}}a_{i+1,\beta}^{\dagger}a_{i+1,\alpha}^{\vphantom{\dagger}}
- a_{i,\beta}^{\dagger}a_{i,\alpha}^{\vphantom{\dagger}}a_{i+1,\alpha}^{\dagger}a_{i+1,\beta}^{\vphantom{\dagger}}\right)&
\intertext{$\bar{n}_i\bar{n}_{i+1}$ is multiplied, where $\bar{n}_i = \sum_{\alpha}\bar{n}_{i,\alpha}$. This does not affect the term as the projected fermions ensure the occupancy is always one or less, and on observing the term, the contribution is non zero only if both sites are occupied.}
&=\sum_{i,\alpha < \beta} \dfrac{2t^2}{U_{\alpha,\beta}}\left(\bar{n}_i^{\vphantom{\dagger}}\bar{n}_{i+1}^{\vphantom{\dagger}}\right) \left(\bar{n}_{i,\alpha}^{\vphantom{\dagger}}\bar{n}_{i+1,\beta}^{\vphantom{\dagger}}+\bar{n}_{i,\beta}^{\vphantom{\dagger}}\bar{n}_{i+1,\alpha}^{\vphantom{\dagger}} \right.&\\
&\phantom{\qquad\qquad\qquad\qquad} \left. -a_{i,\alpha}^{\dagger}a_{i,\beta}^{\vphantom{\dagger}}a_{i+1,\beta}^{\dagger}a_{i+1,\alpha}^{\vphantom{\dagger}} - a_{i,\beta}^{\dagger}a_{i,\alpha}^{\vphantom{\dagger}}a_{i+1,\alpha}^{\dagger}a_{i+1,\beta}^{\vphantom{\dagger}}\right) &
\intertext{The expression is re-written in terms of the spin operators defined in the $\alpha-\beta$ subspace.}
&=\sum_{i,\alpha < \beta} \dfrac{2t^2}{U_{\alpha,\beta}}\left(\bar{n}_i\bar{n}_{i+1}\right)\left(\dfrac{n_{i,(\alpha,\beta)}^{\vphantom{\dagger}}n_{i+1,(\alpha,\beta)}^{\vphantom{\dagger}}-S^z_{i,(\alpha,\beta)}S^z_{i+1,(\alpha,\beta)}}{2} \right.&\\
&\phantom{=\sum_{i,\alpha < \beta} \dfrac{2t^2}{U_{\alpha,\beta}}\left(\bar{n}_i\bar{n}_{i+1}\right)\quad}\left. \dfrac{-S^x_{i,(\alpha,\beta)}S^x_{i+1,(\alpha,\beta)}-S^y_{i,(\alpha,\beta)}S^y_{i+1,(\alpha,\beta)}}{2}\right)&
\intertext{A projection operator ($\hat{P}^{\alpha,\beta}_{i,i+1}$) is introduced that restricts the spins on sites $i$ and $i+1$ to have $\alpha$ or $\beta$ spin components. With the introduction of the projection operator, the spin space reduces to a system of two spin components, and the previously defined spin operators can be written as Pauli matrices ($\sigma^{v}$, $v=x,y,z$). }
&=\sum_{i,\alpha < \beta} \dfrac{2t^2}{U_{\alpha,\beta}}(\bar{n}_i\bar{n}_{i+1})\left(\dfrac{I - \vec{\sigma}_i\vec{\sigma}_{i+1}}{2}\right)\hat{P}^{\alpha,\beta}_{i,i+1}&\\
&=\sum_{i,\alpha < \beta} \dfrac{2t^2}{U_{\alpha,\beta}} (\bar{n}_i\bar{n}_{i+1})(I-\mathcal{E}_{i,i+1})\hat{P}^{\alpha,\beta}_{i,i+1}&
\intertext{In the previous step, the exchange operator ($\mathcal{E}$) replaces the Pauli matrices ($\mathcal{E}_{i,i+1} = \sfrac{I+\vec{\sigma}_i\vec{\sigma}_{i+1}}{2}$). The first process is expressed in a spin-charge separated form.}
&=t^2 \sum_{i} (2\bar{n}_i\bar{n}_{i+1}) \left(\sum_{\alpha<\beta}\dfrac{1}{U_{\alpha,\beta}}(I-\mathcal{E}_{i,i+1})\hat{P}^{\alpha,\beta}_{i,i+1}\right)
\end{flalign*}

Process \text{\textcircled{\scriptsize{2}}} is split into two parts:
\begin{flalign*}
\text{\textcircled{\scriptsize{2}}}: - &\sum_{\expval{i,j,k \neq i},\alpha \neq \beta} \left( \dfrac{t^2}{U_{\alpha,\beta}}(a_{i,\alpha}^{\dagger}\bar{n}_{j,\beta}^{\vphantom{\dagger}}a_{k,\alpha}^{\vphantom{\dagger}}-a_{i,\alpha}^{\dagger}a_{j,\beta}^{\dagger}a_{j,\alpha}^{\vphantom{\dagger}}a_{k,\beta}^{\vphantom{\dagger}}) \right) &\\ 
&= \underbrace{\begin{aligned}-\sum_{i,\alpha < \beta} \dfrac{t^2}{U_{\alpha,\beta}} \Big(&a_{i+2,\alpha}^{\dagger}(\bar{n}_{i+1,\beta}^{\vphantom{\dagger}}a_{i,\alpha}^{\vphantom{\dagger}}-a_{i+1,\beta}^{\dagger}a_{i+1,\alpha}^{\vphantom{\dagger}}a_{i,\beta}^{\vphantom{\dagger}}) +\\ 
& a_{i+2,\beta}^{\dagger}(\bar{n}_{i+1,\alpha}^{\vphantom{\dagger}}a_{i,\beta}^{\vphantom{\dagger}}-a_{i+1,\alpha}^{\dagger}a_{i+1,\beta}^{\vphantom{\dagger}}a_{i,\alpha}^{\vphantom{\dagger}})\Big)\end{aligned}}_{\text{\textcircled{\scriptsize{2a}}}} &\\ 
&\qquad \underbrace{\begin{aligned}-\sum_{i,\alpha < \beta} \dfrac{t^2}{U_{\alpha,\beta}} \Big(&a_{i-1,\alpha}^{\dagger}(\bar{n}_{i,\beta}^{\vphantom{\dagger}}a_{i+1,\alpha}^{\vphantom{\dagger}}-a_{i,\beta}^{\dagger}a_{i,\alpha}^{\vphantom{\dagger}}a_{i+1,\beta}^{\vphantom{\dagger}}) + \\ 
& a_{i-1,\beta}^{\dagger}(\bar{n}_{i,\alpha}^{\vphantom{\dagger}}a_{i+1,\beta}^{\vphantom{\dagger}}-a_{i,\alpha}^{\dagger}a_{i,\beta}^{\vphantom{\dagger}}a_{i+1,\alpha}^{\vphantom{\dagger}})\Big)\end{aligned}}_{\text{\textcircled{\scriptsize{2b}}}}
\end{flalign*}

\begin{flalign*}
&\text{\textcircled{\scriptsize{2a}}}: - \sum_{i,\alpha < \beta} \dfrac{t^2}{U_{\alpha,\beta}}\left(a_{i+2,\alpha}^{\dagger}(\bar{n}_{i+1,\beta}^{\vphantom{\dagger}}a_{i,\alpha}^{\vphantom{\dagger}}-a_{i+1,\beta}^{\dagger}a_{i+1,\alpha}^{\vphantom{\dagger}}a_{i,\beta}^{\vphantom{\dagger}}) + \right.&\\
&\phantom{\qquad\qquad\qquad}\left. a_{i+2,\beta}^{\dagger}(\bar{n}_{i+1,\alpha}^{\vphantom{\dagger}}a_{i,\beta}^{\vphantom{\dagger}}-a_{i+1,\alpha}^{\dagger}a_{i+1,\beta}^{\vphantom{\dagger}}a_{i,\alpha}^{\vphantom{\dagger}})\right)&
\intertext{Adding redundant operators that do not change the outcome: $a_{i,\gamma} \rightarrow a_{i,\gamma}\bar{n}_{i,\gamma}$ and $a_{i,\delta} \rightarrow a_{i,\gamma \neq \delta}a_{i,\gamma \neq \delta}^{\dagger}a_{i,\delta}$. This trick is done to eventually separate the spin and charge degrees of freedom by simple rearrangement.}
&=- \sum_{i,\alpha < \beta} \dfrac{t^2}{U_{\alpha,\beta}} \left(a_{i+2,\alpha}^{\dagger}(\bar{n}_{i+1,\beta}^{\vphantom{\dagger}}a_{i,\alpha}^{\vphantom{\dagger}}\bar{n}_{i,\alpha}^{\vphantom{\dagger}}-a_{i+1,\beta}^{\dagger}a_{i+1,\alpha}^{\vphantom{\dagger}}a_{i,\alpha}^{\vphantom{\dagger}}a_{i,\alpha}^{\dagger}a_{i,\beta}^{\vphantom{\dagger}}) + \right.&\\
&\phantom{\qquad\qquad\qquad}\left. a_{i+2,\beta}^{\dagger}(\bar{n}_{i+1,\alpha}^{\vphantom{\dagger}}a_{i,\beta}^{\vphantom{\dagger}}\bar{n}_{i,\beta}^{\vphantom{\dagger}}-a_{i+1,\alpha}^{\dagger}a_{i+1,\beta}^{\vphantom{\dagger}}a_{i,\beta}^{\vphantom{\dagger}}a_{i,\beta}^{\dagger}a_{i,\alpha}^{\vphantom{\dagger}})\right)&\\
&=- \sum_{i,\alpha < \beta} \dfrac{t^2}{U_{\alpha,\beta}}\left(a_{i+2,\alpha}^{\dagger}\bar{n}_{i+1}^{\vphantom{\dagger}}a_{i,\alpha}^{\vphantom{\dagger}}(\bar{n}_{i+1,\beta}^{\vphantom{\dagger}}\bar{n}_{i,\alpha}^{\vphantom{\dagger}}-a_{i+1,\beta}^{\dagger}a_{i+1,\alpha}^{\vphantom{\dagger}}a_{i,\alpha}^{\dagger}a_{i,\beta}^{\vphantom{\dagger}}) + \right.&\\
&\phantom{\qquad\qquad\qquad}\left. a_{i+2,\beta}^{\dagger}\bar{n}_{i+1}^{\vphantom{\dagger}}a_{i,\beta}^{\vphantom{\dagger}}(\bar{n}_{i+1,\alpha}^{\vphantom{\dagger}}\bar{n}_{i,\beta}^{\vphantom{\dagger}}-a_{i+1,\alpha}^{\dagger}a_{i+1,\beta}^{\vphantom{\dagger}}a_{i,\beta}^{\dagger}a_{i,\alpha}^{\vphantom{\dagger}})\right)&
\intertext{The spin symbols are removed from $a_{i+2,\gamma}^{\dagger}a_{i,\gamma}^{\vphantom{\dagger}} \rightarrow a_{i+2}^{\dagger}a_{i}^{\vphantom{\dagger}}$. This is justified, as the hopping preserves the spin component, and the operators following this in the bracket ensures the spin component information is retained for the hopping.}
&=- \sum_{i,\alpha < \beta} \dfrac{t^2}{U_{\alpha,\beta}} (a_{i+2}^{\dagger}\bar{n}_{i+1}^{\vphantom{\dagger}}a_{i}^{\vphantom{\dagger}})(\bar{n}_{i+1,\beta}^{\vphantom{\dagger}}\bar{n}_{i,\alpha}^{\vphantom{\dagger}}+\bar{n}_{i+1,\alpha}^{\vphantom{\dagger}}\bar{n}_{i,\beta}^{\vphantom{\dagger}} - &\\
&\phantom{\qquad\qquad\qquad\qquad\qquad} a_{i+1,\beta}^{\dagger}a_{i+1,\alpha}^{\vphantom{\dagger}}a_{i,\alpha}^{\dagger}a_{i,\beta}^{\vphantom{\dagger}}-a_{i+1,\alpha}^{\dagger}a_{i+1,\beta}^{\vphantom{\dagger}}a_{i,\beta}^{\dagger}a_{i,\alpha}^{\vphantom{\dagger}})&
\intertext{Finally similar to the derivation in \textcircled{\scriptsize{1}}, the expression is written in terms of spin operators in the $\alpha-\beta$ subspace.}
&=-\sum_{i,\alpha < \beta} \dfrac{t^2}{U_{\alpha,\beta}} (a_{i+2}^{\dagger}\bar{n}_{i+1}^{\vphantom{\dagger}}a_{i}^{\vphantom{\dagger}})\left(\dfrac{n_{i,(\alpha,\beta)}^{\vphantom{\dagger}}n_{i+1,(\alpha,\beta)}^{\vphantom{\dagger}}-S^z_{i,(\alpha,\beta)}S^z_{i+1,(\alpha,\beta)}}{2} \right.&\\
&\phantom{\qquad\qquad\qquad\qquad\qquad}\left. \dfrac{-S^x_{i,(\alpha,\beta)}S^x_{i+1,(\alpha,\beta)}-S^y_{i,(\alpha,\beta)}S^y_{i+1,(\alpha,\beta)}}{2}\right)&
\intertext{This is followed by a projection operator restricting the spin space to $\alpha-\beta$ subspace, and using Pauli matrices to write the expression in terms of exchange operator. The final expression represents a spin-charge separated form.}
&=-t^2 \sum_{i} (a_{i+2}^{\dagger}\bar{n}_{i+1}^{\vphantom{\dagger}}a_{i}^{\vphantom{\dagger}}) \left(\sum_{\alpha<\beta}\dfrac{1}{U_{\alpha,\beta}}(I-\mathcal{E}_{i,i+1})\hat{P}^{\alpha,\beta}_{i,i+1}\right)
\intertext{A similar simplification for part \textcircled{\scriptsize{2b}} is done:}
&\text{\textcircled{\scriptsize{2b}}}: - \sum_{i,\alpha < \beta} \dfrac{t^2}{U_{\alpha,\beta}}\bigg(a_{i-1,\alpha}^{\dagger}(\bar{n}_{i,\beta}^{\vphantom{\dagger}}a_{i+1,\alpha}^{\vphantom{\dagger}} -a_{i,\beta}^{\dagger}a_{i,\alpha}^{\vphantom{\dagger}}a_{i+1,\beta}^{\vphantom{\dagger}}) + &\\
&\phantom{\qquad\qquad\qquad\quad} a_{i-1,\beta}^{\dagger}(\bar{n}_{i,\alpha}^{\vphantom{\dagger}}a_{i+1,\beta}^{\vphantom{\dagger}}-a_{i,\alpha}^{\dagger}a_{i,\beta}^{\vphantom{\dagger}}a_{i+1,\alpha}^{\vphantom{\dagger}})\bigg)&\\
&=- \sum_{i,\alpha < \beta} \dfrac{t^2}{U_{\alpha,\beta}} (a_{i-1}^{\dagger}\bar{n}_{i}^{\vphantom{\dagger}}a_{i+1}^{\vphantom{\dagger}})\bigg(\bar{n}_{i,\beta}^{\vphantom{\dagger}}\bar{n}_{i+1,\alpha}^{\vphantom{\dagger}}+\bar{n}_{i,\alpha}^{\vphantom{\dagger}}\bar{n}_{i+1,\beta}^{\vphantom{\dagger}} - &\\
&\phantom{\qquad\qquad\qquad\quad} a_{i,\beta}^{\dagger}a_{i,\alpha}^{\vphantom{\dagger}}a_{i+1,\alpha}^{\dagger}a_{i+1,\beta}^{\vphantom{\dagger}}-a_{i+1,\alpha}^{\dagger}a_{i+1,\beta}^{\vphantom{\dagger}}a_{i,\beta}^{\dagger}a_{i,\alpha}^{\vphantom{\dagger}}\bigg)
\end{flalign*}
\begin{flalign*}
&=-t^2 \sum_{i} (a_{i-1}^{\dagger}\bar{n}_{i}^{\vphantom{\dagger}}a_{i+1}^{\vphantom{\dagger}}) \left(\sum_{\alpha<\beta}\dfrac{1}{U_{\alpha,\beta}}(I-\mathcal{E}_{i,i+1})\hat{P}^{\alpha,\beta}_{i,i+1}\right) &&
\end{flalign*}
Then, the second term ($\mathcal{Q}^0H\mathcal{Q}^1\frac{1}{\mathcal{Q}^1H\mathcal{Q}^1}\mathcal{Q}^1H\mathcal{Q}^0$) in Eqn.~(\ref{eqn:A1}), in a spin-charge separated form is
\begin{align}
\begin{split}
={}& t^2 \sum_{i} (2\bar{n}_i^{\vphantom{\dagger}}\bar{n}_{i+1}^{\vphantom{\dagger}}-a_{i+2}^{\dagger}\bar{n}_{i+1}^{\vphantom{\dagger}}a_{i}^{\vphantom{\dagger}}-a_{i-1}^{\dagger}\bar{n}_{i}^{\vphantom{\dagger}}a_{i+1}^{\vphantom{\dagger}}) \left(\sum_{\alpha<\beta}\dfrac{(I-\mathcal{E}_{i,i+1})}{U_{\alpha,\beta}}\hat{P}^{\alpha,\beta}_{i,i+1}\right)
\end{split}
\end{align}
The effective Hamiltonian after second order expansion, followed by separation of spin \& charge degrees of freedom is 
\begin{widetext}
\begin{equation}
H_{\text{eff}} = -t\sum_{i}(a_{i}^{\dagger}a_{i+1}^{\vphantom{\dagger}}+a_{i+1}^{\dagger}a_{i}^{\vphantom{\dagger}}) -t^2 \sum_{i} (2\bar{n}_i^{\vphantom{\dagger}}\bar{n}_{i+1}^{\vphantom{\dagger}}-a_{i+2}^{\dagger}\bar{n}_{i+1}^{\vphantom{\dagger}}a_{i}^{\vphantom{\dagger}}-a_{i-1}^{\dagger}\bar{n}_{i}^{\vphantom{\dagger}}a_{i+1}^{\vphantom{\dagger}}) \left(\sum_{\alpha<\beta}\dfrac{1}{U_{\alpha,\beta}}(I-\mathcal{E}_{i,i+1})\hat{P}^{\alpha,\beta}_{i,i+1}\right)\,.
\end{equation}
Treat the effective Hamiltonian as an unperturbed Hamiltonian with a perturbation. The unperturbed part is the tunneling of the spinless fermions. Using the wavefunction in spatial sector 1:
\begin{align} 
    \Psi(i_1,\sigma_1...,i_{M},\sigma_{M})=\varphi(i_1,i_2..i_{M})\chi\,,
\end{align}
and applying degenerate perturbation theory, we obtain the spin-chain Hamiltonian for fermions ($\expval{H}{\varphi}$): 
\begin{align}
\begin{split}
H_{\text{SC}}  &= \expval{-t\sum_{i}(a_{i}^{\dagger}a_{i+1}^{\vphantom{\dagger}}+a_{i+1}^{\dagger}a_{i}^{\vphantom{\dagger}})}{\varphi} 
+ \expval{-t^2\sum_{i} (2\bar{n}_i^{\vphantom{\dagger}}\bar{n}_{i+1}^{\vphantom{\dagger}}-a_{i+2}^{\dagger}\bar{n}_{i+1}^{\vphantom{\dagger}}a_{i}^{\vphantom{\dagger}}-a_{i-1}^{\dagger}\bar{n}_{i}^{\vphantom{\dagger}}a_{i+1}^{\vphantom{\dagger}}) \left(\sum_{\alpha<\beta}\dfrac{1}{U_{\alpha,\beta}}(I-\mathcal{E}_{i,i+1})\hat{P}^{\alpha,\beta}_{i,i+1}\right)}{\varphi}
\\ &= E_0- \sum_{l}t^2\expval{\sum_i \delta_{\sum_{j=1}^{i-1}\bar{n}_j,l-1 }(2\bar{n}_i^{\vphantom{\dagger}}\bar{n}_{i+1}^{\vphantom{\dagger}}-a_{i+2}^{\dagger}\bar{n}_{i+1}^{\vphantom{\dagger}}a_{i}^{\vphantom{\dagger}}-a_{i-1}^{\dagger}\bar{n}_{i}^{\vphantom{\dagger}}a_{i+1}^{\vphantom{\dagger}})}{\varphi}  \otimes\left(\sum_{\alpha<\beta}\dfrac{1}{U_{\alpha,\beta}}(I-\mathcal{E}_{l,l+1})\hat{P}^{\alpha,\beta}_{l,l+1}\right)
\\ &= E_0- t^2\sum_{l}\mathcal{C}_l \left(\sum_{\alpha<\beta}\dfrac{1}{U_{\alpha,\beta}}(I-\mathcal{E}_{l,l+1})\hat{P}^{\alpha,\beta}_{l,l+1}\right),
\end{split}
\end{align}
where $E_0$ is the ground state energy of the unperturbed Hamiltonian, the tunneling of spinless fermions. A delta function ($\delta_{\sum_{j=1}^{i-1}\bar{n}_j,l-1 }$) is introduced, in the expectation of the charge operators to ensure the charge operators acting on sites $i$ and $i+1$ correspond to the spin operators on spins $l$ and $l+1$. $C_l$ is the coefficient of the operator acting on the spin wavefunction, containing all the charge degrees of freedom. This provides the effective spin-chain for $N$-component Fermi--Hubbard Hamiltonian.

\section{\label{sec:appendB}Effective Spin-Chain Hamiltonian: Bosons}
\begin{flalign}
\text{As for fermions, the zeroth term in the effective Hamiltonian (Eqn.~\ref{eqn:A1}):~} \mathcal{Q}^0 H\mathcal{Q}^0 = -t\sum_{i,\alpha}\mathcal{Q}^0(b_{i,\alpha}^{\dagger}b_{i+1,\alpha}^{\vphantom{\dagger}}+b_{i+1,\alpha}^{\dagger}b_{i,\alpha}^{\vphantom{\dagger}})\mathcal{Q}^0\,. &&
\end{flalign}
\noindent For the second order term: 
\begin{align*}
\begin{split}
\mathcal{Q}^0 H\mathcal{Q}^1&\dfrac{1}{\mathcal{Q}^1 H\mathcal{Q}^1}\mathcal{Q}^1H\mathcal{Q}^0  
\\ = &\sum_{\expval{i,j,k},\alpha,\beta}t^2\mathcal{Q}^0(b_{i,\alpha}^{\dagger}b_{j,\alpha}^{\vphantom{\dagger}})\dfrac{\mathcal{Q}^1}{\mathcal{Q}^1 H\mathcal{Q}^1}(b_{j,\beta}^{\dagger}b_{k,\beta}^{\vphantom{\dagger}})\mathcal{Q}^0 
\\ = &\sum_{\expval{i,j,k},\alpha=\beta} \left[ t^2\mathcal{Q}^0(b_{i,\alpha}^{\dagger}b_{j,\alpha}^{\vphantom{\dagger}})\dfrac{(n_{j,\alpha}-1)}{V_{\alpha}}(b_{j,\alpha}^{\dagger}b_{k,\alpha}^{\vphantom{\dagger}})\mathcal{Q}^0 +
t^2\mathcal{Q}^0(b_{i,\alpha}^{\dagger}b_{j,\alpha}^{\vphantom{\dagger}})\dfrac{\sum_{\gamma \neq \alpha} n_{j,\gamma}}{U_{\alpha,\gamma}}(b_{j,\alpha}^{\dagger}b_{k,\alpha}^{\vphantom{\dagger}})\mathcal{Q}^0 \right] +
\sum_{\expval{i,j,k},\alpha \neq \beta} t^2\mathcal{Q}^0 (b_{i,\alpha}^{\dagger}b_{j,\alpha}^{\vphantom{\dagger}})\dfrac{1}{U_{\alpha,\beta}}(b_{j,\beta}^{\dagger}b_{k,\beta}^{\vphantom{\dagger}})\mathcal{Q}^0
\\ = &\phantom{\beta}\sum_{\expval{i,j,k},\alpha}2t^2\mathcal{Q}^0(b_{i,\alpha}^{\dagger}\dfrac{n_{j,\alpha}}{V_{\alpha}}b_{k,\alpha}^{\vphantom{\dagger}})\mathcal{Q}^0 +
\sum_{\expval{i,j,k},\alpha \neq \gamma}t^2\mathcal{Q}^0(b_{i,\alpha}^{\dagger}\dfrac{n_{j,\gamma}}{U_{\alpha,\gamma}}b_{k,\alpha}^{\vphantom{\dagger}})\mathcal{Q}^0 + \sum_{\expval{i,j,k},\alpha \neq \beta} t^2\mathcal{Q}^0 (b_{i,\alpha}^{\dagger}\dfrac{b_{j,\beta}^{\dagger}b_{j,\alpha}^{\vphantom{\dagger}}}{U_{\alpha,\beta}}b_{k,\beta}^{\vphantom{\dagger}})\mathcal{Q}^0\,.
\end{split}
\end{align*}
Hence, the Bosonic effective Hamiltonian
\begin{align}
\begin{split}
H^{\text{B}}_{\text{eff}} =  -t&\sum_{i,\alpha}\mathcal{Q}^0\left(b_{i,\alpha}^{\dagger}b_{i+1,\alpha}^{\vphantom{\dagger}}+b_{i+1,\alpha}^{\dagger}b_{i,\alpha}^{\vphantom{\dagger}}\right)\mathcal{Q}^0  
- \sum_{i,\alpha}4t^2\mathcal{Q}^0\left(\dfrac{n_{i,\alpha}n_{i+1,\alpha}}{V_{\alpha}}\right)\mathcal{Q}^0
-\sum_{i,\alpha}2t^2\mathcal{Q}^0\left(b_{i+2,\alpha}^{\dagger}\dfrac{n_{i+1,\alpha}}{V_{\alpha}}b_{i,\alpha}^{\vphantom{\dagger}} + b_{i-1,\alpha}^{\dagger}\dfrac{n_{i,\alpha}}{V_{\alpha}}b_{i+1,\alpha}^{\vphantom{\dagger}}\right)\mathcal{Q}^0 -
\\ & \sum_{i,\alpha < \beta}2t^2\mathcal{Q}^0\left(\dfrac{n_{i,\alpha}n_{i+1,\beta}+n_{i,\beta}n_{i+1,\alpha}}{U_{\alpha,\beta}}\right)\mathcal{Q}^0
-\sum_{i,\alpha<\beta}t^2\mathcal{Q}^0\left(b_{i+2,\alpha}^{\dagger}\dfrac{n_{i+1,\beta}}{U_{\alpha,\beta}}b_{i,\alpha}^{\vphantom{\dagger}} + b_{i+2,\beta}^{\dagger}\dfrac{n_{i+1,\alpha}}{U_{\alpha,\beta}}b_{i,\beta}^{\vphantom{\dagger}} + b_{i-1,\alpha}^{\dagger}\dfrac{n_{i,\beta}}{U_{\alpha,\beta}}b_{i+1,\alpha}^{\vphantom{\dagger}} + b_{i-1,\beta}^{\dagger}\dfrac{n_{i,\alpha}}{U_{\alpha,\beta}}b_{i+1,\beta}^{\vphantom{\dagger}}\right)\mathcal{Q}^0 -
\\ & \sum_{i,\alpha < \beta} 2t^2\mathcal{Q}^0 \left(\dfrac{b_{i,\alpha}^{\dagger}b_{i,\beta}^{\vphantom{\dagger}}b_{i+1,\beta}^{\dagger}b_{i+1,\alpha}^{\vphantom{\dagger}}+b_{i,\beta}^{\dagger}b_{i,\alpha}^{\vphantom{\dagger}}b_{i+1,\alpha}^{\dagger}b_{i+1,\beta}^{\vphantom{\dagger}}}{U_{\alpha,\beta}}\right)\mathcal{Q}^0 -
\\ &\sum_{i,\alpha<\beta}t^2\mathcal{Q}^0\left(b_{i+2,\alpha}^{\dagger}\dfrac{b_{i+1,\beta}^{\dagger}b_{i+1,\alpha}^{\vphantom{\dagger}}}{U_{\alpha,\beta}}b_{i,\beta}^{\vphantom{\dagger}}  + b_{i+2,\beta}^{\dagger}\dfrac{b_{i+1,\alpha}^{\dagger}b_{i+1,\beta}^{\vphantom{\dagger}}}{U_{\alpha,\beta}}b_{i,\alpha}^{\vphantom{\dagger}} + b_{i-1,\alpha}^{\dagger}\dfrac{b_{i,\beta}^{\dagger}b_{i,\alpha}^{\vphantom{\dagger}}}{U_{\alpha,\beta}}b_{i+1,\beta}^{\vphantom{\dagger}} + b_{i-1,\beta}^{\dagger}\dfrac{b_{i,\alpha}^{\dagger}b_{i,\beta}^{\vphantom{\dagger}}}{U_{\alpha,\beta}}b_{i+1,\alpha}^{\vphantom{\dagger}}\right)\mathcal{Q}^0\,.
\end{split}
\end{align}
The bosonic effective Hamiltonian contains the projection operator $\mathcal{Q}^0$. To project to low energy subspace, remove all double occupancy by first mapping to fermions (via Jordan Wigner transformation) which prevents higher occupancy of same component bosons. This is followed by mapping the Jordan Wigner fermions to projected fermions. 

The Jordan Wigner transformation for $N$-component bosons (ensuring the bosonic operators follow hard-core boson commutations and fermions anti-commute) is
\begin{align}
    b_{i,\alpha}^{\dagger} = \exp(-i\pi\sum_{l<i}n_l) \exp(-i\pi\sum_{\gamma<\alpha}n_{i,\gamma}) c_{i,\alpha}^{\dagger}\,,
\end{align}
where $n_l = \sum_{\alpha}b_{l,\alpha}^{\dagger}b_{l,\alpha} = \sum_{\alpha}c_{l,\alpha}^{\dagger}c_{l,\alpha}$.
Similar to fermions, the Jordan Wigner fermions are mapped to projected fermions. 

After mapping bosons to fermions and constraining the subspace to occupancy to $\leq1$, the fermionic effective Hamiltonian
\begin{align}
\begin{split}
H^{B\rightarrow F}_{\text{eff}} = -t&\sum_{i,\alpha}(a_{i,\alpha}^{\dagger}a_{i+1,\alpha}+a_{i+1,\alpha}^{\dagger}a_{i,\alpha})
- \sum_{i,\alpha}4t^2(\dfrac{\bar{n}_{i,\alpha}\bar{n}_{i+1,\alpha}}{V_{\alpha}})
+\sum_{i,\alpha}2t^2(a_{i+2,\alpha}^{\dagger}\dfrac{\bar{n}_{i+1,\alpha}}{V_{\alpha}}a_{i,\alpha} + a_{i-1,\alpha}^{\dagger}\dfrac{\bar{n}_{i,\alpha}}{V_{\alpha}}a_{i+1,\alpha})
- \sum_{i,\alpha < \beta}2t^2(\dfrac{\bar{n}_{i,\alpha}\bar{n}_{i+1,\beta}+\bar{n}_{i,\beta}\bar{n}_{i+1,\alpha}}{U_{\alpha,\beta}}) +
\\ &\sum_{i,\alpha<\beta}t^2(a_{i+2,\alpha}^{\dagger}\dfrac{\bar{n}_{i+1,\beta}}{U_{\alpha,\beta}}a_{i,\alpha} + a_{i+2,\beta}^{\dagger}\dfrac{\bar{n}_{i+1,\alpha}}{U_{\alpha,\beta}}a_{i,\beta} + a_{i-1,\alpha}^{\dagger}\dfrac{\bar{n}_{i,\beta}}{U_{\alpha,\beta}}a_{i+1,\alpha} + a_{i-1,\beta}^{\dagger}\dfrac{\bar{n}_{i,\alpha}}{U_{\alpha,\beta}}a_{i+1,\beta}) -
\\ &\sum_{i,\alpha < \beta} 2t^2 (\dfrac{a_{i,\alpha}^{\dagger}a_{i,\beta}a_{i+1,\beta}^{\dagger}a_{i+1,\alpha}+a_{i,\beta}^{\dagger}a_{i,\alpha}a_{i+1,\alpha}^{\dagger}a_{i+1,\beta}}{U_{\alpha,\beta}}) +
\\ &\sum_{i,\alpha<\beta}t^2(a_{i+2,\alpha}^{\dagger}\dfrac{a_{i+1,\beta}^{\dagger}a_{i+1,\alpha}}{U_{\alpha,\beta}}a_{i,\beta} + a_{i+2,\beta}^{\dagger}\dfrac{a_{i+1,\alpha}^{\dagger}a_{i+1,\beta}}{U_{\alpha,\beta}}a_{i,\alpha} + a_{i-1,\alpha}^{\dagger}\dfrac{a_{i,\beta}^{\dagger}a_{i,\alpha}}{U_{\alpha,\beta}}a_{i+1,\beta} + a_{i-1,\beta}^{\dagger}\dfrac{a_{i,\alpha}^{\dagger}a_{i,\beta}}{U_{\alpha,\beta}}a_{i+1,\alpha})\,.
\end{split}
\end{align}
To simplify the expression, define $\mu_{\alpha;\beta} = \sfrac{U_{\alpha,\beta}}{(N-1)V_{\alpha}}$ and rewrite the expression:
\begin{align*}
\begin{split}
H^{B\rightarrow F}_{\text{eff}} = -t&\sum_{i,\alpha}\left(a_{i,\alpha}^{\dagger}a_{i+1,\alpha}^{\vphantom{\dagger}}+a_{i+1,\alpha}^{\dagger}a_{i,\alpha}^{\vphantom{\dagger}}\right)
- \sum_{i,\alpha<\beta}4t^2\left(\dfrac{\mu_{\alpha;\beta}\bar{n}_{i,\alpha}\bar{n}_{i+1,\alpha} + \mu_{\beta;\alpha}\bar{n}_{i,\beta}\bar{n}_{i+1,\beta}}{U_{\alpha,\beta}}\right) +
\\ &\sum_{i,\alpha<\beta}2t^2\left(\mu_{\alpha;\beta}^{\vphantom{\dagger}}a_{i+2,\alpha}^{\dagger}\dfrac{\bar{n}_{i+1,\alpha}}{U_{\alpha,\beta}}a_{i,\alpha} + \mu_{\beta;\alpha}^{\vphantom{\dagger}}a_{i+2,\beta}^{\dagger}\dfrac{\bar{n}_{i+1,\beta}}{U_{\alpha,\beta}}a_{i,\beta} + \mu_{\alpha;\beta}^{\vphantom{\dagger}}a_{i-1,\alpha}^{\dagger}\dfrac{\bar{n}_{i,\alpha}}{U_{\alpha,\beta}}a_{i+1,\alpha} + \mu_{\beta;\alpha}^{\vphantom{\dagger}}a_{i-1,\beta}^{\dagger}\dfrac{\bar{n}_{i,\beta}}{U_{\alpha,\beta}}a_{i+1,\beta}\right) -
\\ &\sum_{i,\alpha < \beta}2t^2\left(\dfrac{2\bar{n}_{i,\alpha}\bar{n}_{i+1,\alpha}+ 2\bar{n}_{i,\beta}\bar{n}_{i+1,\beta}+ \bar{n}_{i,\alpha}\bar{n}_{i+1,\beta}+\bar{n}_{i,\beta}\bar{n}_{i+1,\alpha}}{U_{\alpha,\beta}}\right)
+ \sum_{i,\alpha < \beta}4t^2\left(\dfrac{\bar{n}_{i,\alpha}\bar{n}_{i+1,\alpha}+\bar{n}_{i,\beta}\bar{n}_{i+1,\beta}}{U_{\alpha,\beta}}\right) +
\\ & \sum_{i,\alpha<\beta}t^2\left(a_{i+2,\alpha}^{\dagger}\dfrac{(2\bar{n}_{i+1,\alpha}+\bar{n}_{i+1,\beta})}{U_{\alpha,\beta}}a_{i,\alpha}^{\vphantom{\dagger}} + a_{i+2,\beta}^{\dagger}\dfrac{(2\bar{n}_{i+1,\beta}+\bar{n}_{i+1,\alpha})}{U_{\alpha,\beta}}a_{i,\beta}^{\vphantom{\dagger}}\right)
-\sum_{i,\alpha<\beta}2t^2\left(a_{i+2,\alpha}^{\dagger}\dfrac{\bar{n}_{i+1,\alpha}}{U_{\alpha,\beta}}a_{i,\alpha}^{\vphantom{\dagger}} + a_{i+2,\beta}^{\dagger}\dfrac{\bar{n}_{i+1,\beta}}{U_{\alpha,\beta}}a_{i,\beta}^{\vphantom{\dagger}}\right) +
\\ &\sum_{i,\alpha<\beta}t^2\left(a_{i-1,\alpha}^{\dagger}\dfrac{(2\bar{n}_{i,\alpha}+\bar{n}_{i,\beta})}{U_{\alpha,\beta}}a_{i+1,\alpha}^{\vphantom{\dagger}} + a_{i-1,\beta}^{\dagger}\dfrac{(2\bar{n}_{i,\beta}+\bar{n}_{i,\alpha})}{U_{\alpha,\beta}}a_{i+1,\beta}^{\vphantom{\dagger}}\right)
-\sum_{i,\alpha<\beta}2t^2\left(a_{i-1,\alpha}^{\dagger}\dfrac{\bar{n}_{i,\alpha}}{U_{\alpha,\beta}}a_{i+1,\alpha}^{\vphantom{\dagger}} + a_{i-1,\beta}^{\dagger}\dfrac{\bar{n}_{i,\beta}}{U_{\alpha,\beta}}a_{i+1,\beta}^{\vphantom{\dagger}}\right) -
\\ &\sum_{i,\alpha < \beta} 2t^2\left(\dfrac{a_{i,\alpha}^{\dagger}a_{i,\beta}^{\vphantom{\dagger}}a_{i+1,\beta}^{\dagger}a_{i+1,\alpha}^{\vphantom{\dagger}}+a_{i,\beta}^{\dagger}a_{i,\alpha}^{\vphantom{\dagger}}a_{i+1,\alpha}^{\dagger}a_{i+1,\beta}^{\vphantom{\dagger}}}{U_{\alpha,\beta}}\right)+
\\ & \sum_{i,\alpha<\beta}t^2\left(a_{i+2,\alpha}^{\dagger}\dfrac{a_{i+1,\beta}^{\dagger}a_{i+1,\alpha}^{\vphantom{\dagger}}}{U_{\alpha,\beta}}a_{i,\beta}^{\vphantom{\dagger}} + a_{i+2,\beta}^{\dagger}\dfrac{a_{i+1,\alpha}^{\dagger}a_{i+1,\beta}^{\vphantom{\dagger}}}{U_{\alpha,\beta}}a_{i,\alpha}^{\vphantom{\dagger}} + a_{i-1,\alpha}^{\dagger}\dfrac{a_{i,\beta}^{\dagger}a_{i,\alpha}^{\vphantom{\dagger}}}{U_{\alpha,\beta}}a_{i+1,\beta}^{\vphantom{\dagger}} + a_{i-1,\beta}^{\dagger}\dfrac{a_{i,\alpha}^{\dagger}a_{i,\beta}^{\vphantom{\dagger}}}{U_{\alpha,\beta}}a_{i+1,\alpha}^{\vphantom{\dagger}}\right)
\end{split}
\end{align*}
Group together terms for simplification:
\begin{align*}
\begin{split}
H^{B\rightarrow F}_{\text{eff}} = -t&\sum_{i,\alpha}\left(a_{i,\alpha}^{\dagger}a_{i+1,\alpha}^{\vphantom{\dagger}}+a_{i+1,\alpha}^{\dagger}a_{i,\alpha}^{\vphantom{\dagger}}\right)
- \sum_{i,\alpha<\beta}4t^2\left(\dfrac{(\mu_{\alpha;\beta}-1)\bar{n}_{i,\alpha}\bar{n}_{i+1,\alpha} + (\mu_{\beta;\alpha}-1)\bar{n}_{i,\beta}\bar{n}_{i+1,\beta}}{U_{\alpha,\beta}}\right)+
\\& \sum_{i,\alpha<\beta}2t^2\left((\mu_{\alpha;\beta}-1)a_{i+2,\alpha}^{\dagger}\dfrac{\bar{n}_{i+1,\alpha}}{U_{\alpha,\beta}}a_{i,\alpha}^{\vphantom{\dagger}} + (\mu_{\beta;\alpha}-1)a_{i+2,\beta}^{\dagger}\dfrac{\bar{n}_{i+1,\beta}}{U_{\alpha,\beta}}a_{i,\beta}^{\vphantom{\dagger}}\right)+ 
\\ & \sum_{i,\alpha<\beta}2t^2\left((\mu_{\alpha;\beta}-1)a_{i-1,\alpha}^{\dagger}\dfrac{\bar{n}_{i,\alpha}^{\vphantom{\dagger}}}{U_{\alpha,\beta}}a_{i+1,\alpha}^{\vphantom{\dagger}} + (\mu_{\beta;\alpha}-1)a_{i-1,\beta}^{\dagger}\dfrac{\bar{n}_{i,\beta}^{\vphantom{\dagger}}}{U_{\alpha,\beta}}a_{i+1,\beta}^{\vphantom{\dagger}}\right)-
\\ & \sum_{i,\alpha < \beta}2t^2\left(\dfrac{2\bar{n}_{i,\alpha}^{\vphantom{\dagger}}\bar{n}_{i+1,\alpha}^{\vphantom{\dagger}}+ 2\bar{n}_{i,\beta}^{\vphantom{\dagger}}\bar{n}_{i+1,\beta}^{\vphantom{\dagger}}+ \bar{n}_{i,\alpha}\bar{n}_{i+1,\beta}^{\vphantom{\dagger}}+\bar{n}_{i,\beta}\bar{n}_{i+1,\alpha}^{\vphantom{\dagger}}}{U_{\alpha,\beta}} +\dfrac{a_{i,\alpha}^{\dagger}a_{i,\beta}^{\vphantom{\dagger}}a_{i+1,\beta}^{\dagger}a_{i+1,\alpha}^{\vphantom{\dagger}}+a_{i,\beta}^{\dagger}a_{i,\alpha}^{\vphantom{\dagger}}a_{i+1,\alpha}^{\dagger}a_{i+1,\beta}^{\vphantom{\dagger}}}{U_{\alpha,\beta}}\right)+ 
\\ & 
\sum_{i,\alpha<\beta}t^2\left(\dfrac{a_{i+2,\alpha}^{\dagger}}{U_{\alpha,\beta}}(2\bar{n}_{i+1,\alpha}^{\vphantom{\dagger}}a_{i,\alpha}^{\vphantom{\dagger}} +\bar{n}_{i+1,\beta}^{\vphantom{\dagger}}a_{i,\alpha}^{\vphantom{\dagger}} +a_{i+1,\alpha}^{\dagger}a_{i+1,\beta}^{\vphantom{\dagger}}a_{i,\beta}^{\vphantom{\dagger}} )
+ \dfrac{a_{i+2,\beta}^{\dagger}}{U_{\alpha,\beta}}(2\bar{n}_{i+1,\beta}^{\vphantom{\dagger}}a_{i,\beta}^{\vphantom{\dagger}} +\bar{n}_{i+1,\alpha}^{\vphantom{\dagger}}a_{i,\beta}^{\vphantom{\dagger}} +a_{i+1,\beta}^{\dagger}a_{i+1,\alpha}^{\vphantom{\dagger}}a_{i,\alpha}^{\vphantom{\dagger}} ) \right)+ 
\\ & \sum_{i,\alpha<\beta}t^2\left(\dfrac{a_{i-1,\alpha}^{\dagger}}{U_{\alpha,\beta}}(2\bar{n}_{i,\alpha}^{\vphantom{\dagger}}a_{i+1,\alpha}^{\vphantom{\dagger}} +\bar{n}_{i,\beta}^{\vphantom{\dagger}}a_{i+1,\alpha}^{\vphantom{\dagger}} +a_{i,\alpha}^{\dagger}a_{i,\beta}^{\vphantom{\dagger}}a_{i+1,\beta}^{\vphantom{\dagger}} )
+ \dfrac{a_{i-1,\beta}^{\dagger}}{U_{\alpha,\beta}}(2\bar{n}_{i,\beta}^{\vphantom{\dagger}}a_{i+1,\beta}^{\vphantom{\dagger}} +\bar{n}_{i,\alpha}^{\vphantom{\dagger}}a_{i+1,\beta}^{\vphantom{\dagger}} +a_{i,\beta}^{\dagger}a_{i,\alpha}^{\vphantom{\dagger}}a_{i+1,\alpha}^{\vphantom{\dagger}} ) \right) 
\end{split}
\end{align*}
Using the spin operators (Eqn.~\ref{eqn:A4}), as for fermions, the effective Hamiltonian for $N$-component bosons is simplified further:
\begin{align*}
\begin{split}
H^{B\rightarrow F}_{\text{eff}} =  -t&\sum_{i,\alpha}(a_{i,\alpha}^{\dagger}a_{i+1,\alpha}^{\vphantom{\dagger}}+a_{i+1,\alpha}^{\dagger}a_{i,\alpha}^{\vphantom{\dagger}}) -
\sum_{i,\alpha<\beta}t^2\dfrac{(2\bar{n}_i\bar{n}_{i+1})}{U_{\alpha,\beta}}\Bigg[(\dfrac{\mu_{\alpha;\beta}+\mu_{\beta;\alpha}}{2}-1)(I+S_{i,(\alpha,\beta)}^zS_{i+1,(\alpha,\beta)}^z) + \dfrac{\mu_{\beta;\alpha}-\mu_{\alpha;\beta}}{2}(S_{i,(\alpha,\beta)}^z+S_{i+1,(\alpha,\beta)}^z)\Bigg]\hat{P}_{i,i+1}^{\alpha,\beta} +
\\ &\sum_{i,\alpha<\beta}t^2(a_{i+2}^{\dagger}\bar{n}_{i+1}^{\vphantom{\dagger}}a_i^{\vphantom{\dagger}}+a_{i-1}^{\dagger}\bar{n}_i^{\vphantom{\dagger}}a_{i+1}^{\vphantom{\dagger}})\dfrac{1}{U_{\alpha,\beta}}\Bigg[\left(\dfrac{\mu_{\alpha;\beta}+\mu_{\beta;\alpha}}{2}-1\right)(I+S_{i,(\alpha,\beta)}^zS_{i+1,(\alpha,\beta)}^z) + \dfrac{\mu_{\beta;\alpha}-\mu_{\alpha;\beta}}{2}(S_{i,(\alpha,\beta)}^z+S_{i+1,(\alpha,\beta)}^z)\Bigg]\hat{P}_{i,i+1}^{\alpha,\beta} -
\\ & \sum_{i,\alpha < \beta}t^2(2\bar{n}_i\bar{n}_{i+1})\dfrac{1}{U_{\alpha,\beta}}(I + \mathcal{E}_{i,i+1})\hat{P}_{i,i+1}^{\alpha,\beta} +
\sum_{i,\alpha<\beta}t^2(a_{i+2}^{\dagger}\bar{n}_{i+1}^{\vphantom{\dagger}}a_i^{\vphantom{\dagger}}+a_{i-1}^{\dagger}\bar{n}_i^{\vphantom{\dagger}}a_{i+1}^{\vphantom{\dagger}})\dfrac{1}{U_{\alpha,\beta}}(I + \mathcal{E}_{i,i+1})\hat{P}_{i,i+1}^{\alpha,\beta}\\
H^{B\rightarrow F}_{\text{eff}} =  -t&\sum_{i}(a_{i}^{\dagger}a_{i+1}^{\vphantom{\dagger}}+a_{i+1}^{\dagger}a_{i}^{\vphantom{\dagger}}) -
\sum_{i,\alpha < \beta}t^2(2\bar{n}_i^{\vphantom{\dagger}}\bar{n}_{i+1}^{\vphantom{\dagger}}-a_{i+2}^{\dagger}\bar{n}_{i+1}^{\vphantom{\dagger}}a_i^{\vphantom{\dagger}}-a_{i-1}^{\dagger}\bar{n}_i^{\vphantom{\dagger}}a_{i+1}^{\vphantom{\dagger}})\dfrac{1}{U_{\alpha,\beta}}(I + \mathcal{E}_{i,i+1})\hat{P}_{i,i+1}^{\alpha,\beta}
- \\
&\sum_{i,\alpha<\beta}t^2(2\bar{n}_i^{\vphantom{\dagger}}\bar{n}_{i+1}^{\vphantom{\dagger}}-a_{i+2}^{\dagger}\bar{n}_{i+1}^{\vphantom{\dagger}}a_i^{\vphantom{\dagger}}-a_{i-1}^{\dagger}\bar{n}_i^{\vphantom{\dagger}}a_{i+1}^{\vphantom{\dagger}}) \dfrac{1}{U_{\alpha,\beta}}\bigg[\left(\dfrac{\mu_{\alpha;\beta}+\mu_{\beta;\alpha}}{2}-1\right)
(I+S_{i,(\alpha,\beta)}^zS_{i+1,(\alpha,\beta)}^z) + \dfrac{\mu_{\beta;\alpha}-\mu_{\alpha;\beta}}{2}(S_{i,(\alpha,\beta)}^z+S_{i+1,(\alpha,\beta)}^z)\bigg]\hat{P}_{i,i+1}^{\alpha,\beta} 
\end{split}
\end{align*}
The zeroth order term, preserves spin while hopping and is summed over all spins. Hence, the spin symbol is removed from the zeroth order term. Similar to fermions, obtaining an effective spin-chain Hamiltonian for $N$-component bosons as
\begin{align}
H_{\text{SC}}^{\text{B}} =  E_0- t^2\sum_{l}\mathcal{C}_l\left[\sum_{\alpha<\beta}\dfrac{1}{U_{\alpha,\beta}}\left[(I+\mathcal{E}_{l,l+1}) + (\dfrac{\mu_{\alpha;\beta}+\mu_{\beta;\alpha}}{2}-1)(I+S_{l,(\alpha,\beta)}^zS_{l+1,(\alpha,\beta)}^z) + \dfrac{\mu_{\beta;\alpha}-\mu_{\alpha;\beta}}{2}(S_{l,(\alpha,\beta)}^z+S_{l+1,(\alpha,\beta)}^z) \right] \hat{P}^{\alpha,\beta}_{l,l+1}\right], \\
\text{where the coefficient~} \mathcal{C}_l = \expval{\sum_{j=1}^{L-1} \delta_{\sum_{j=1}^{i-1}\bar{n}_j,l-1}(2\bar{n}_i^{\vphantom{\dagger}}\bar{n}_{i+1}^{\vphantom{\dagger}}-a_{i+2}^{\dagger}\bar{n}_{i+1}^{\vphantom{\dagger}}a_{i}^{\vphantom{\dagger}}-a_{i-1}^{\dagger}\bar{n}_{i}^{\vphantom{\dagger}}a_{i+1}^{\vphantom{\dagger}})}{\varphi}\,.
\end{align}
\end{widetext}

\section{\label{sec:appendC}Spin-1/2 Spinor Gas}

\begin{figure*}[b]
\centering
\includegraphics{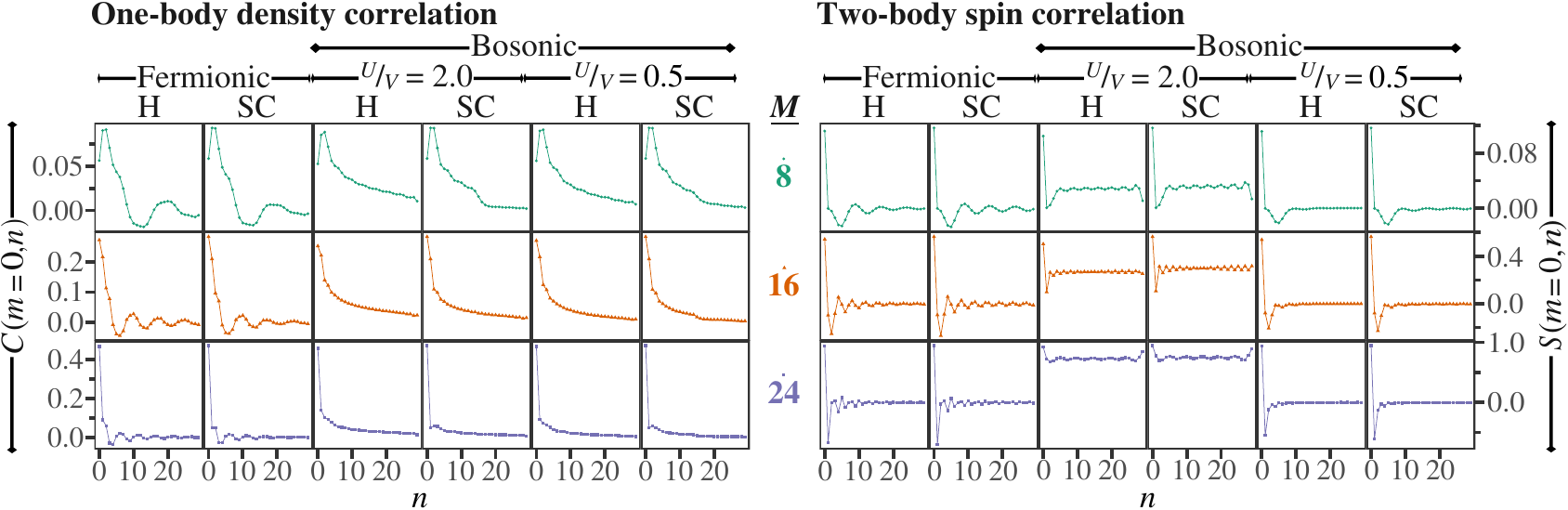}
\caption{\label{fig:append}(color online) One-body density correlation ($C(m=0,n)$ and diagonal two-body spin correlation ($S(m=0,n)$) as a function of site $n$ for fermions and bosons with inter-to intra-component interaction strength $\sfrac{U}{V}=\{2, 0.5\}$ using the Hubbard Hamiltonian (H) and spin-chain Hamiltonian (SC). The correlations are obtained for number of particles $M= \{8, 16, 24\}$, system size $L=30$, open boundary conditions and in the strong interaction region ($\sfrac{U}{t}=40$).}
\end{figure*}

Here, we study the spin-$\sfrac{1}{2}$ spinor gas trapped in a 1D optical lattice at strong on-site interaction. Fig.~\ref{fig:append} provides a comparison of the one-body density correlation and diagonal two-body spin correlations of fermions and bosons ($\sfrac{U}{V}=2$, $0.5$) obtained via the generalized effective spin-chain Hamiltonian (SC) acting on the mapped system to the Hubbard Hamiltonian (H) on the original system. The number of particles are $M=8$, $16$, $24$ for number of lattice sites $L=30$ and open boundary conditions. The Hilbert space of the $2-$component Bose--Hubbard Hamiltonian is limited to at maximum occupancy $2$ per site. This restriction is justified for the considered on-site interaction strengths and confirmed with no double occupancy in the ground state. In the one-body density correlation $(C(m=0,n)$, at small occupancy ($\sfrac{M}{L}$), the correlations peak for nearest neighboring sites. However, as the occupancy increases, the maxima occurs when $m=n=0$. This holds true for both fermions and bosons ($\sfrac{U}{V}=2$, $0.5$). The $k_\text{F}$ oscillations are displayed for fermions but not for bosons. Additionally, for bosons irrespective of the $\sfrac{U}{V}$ consideration, the correlations are identical. The antiferromagnetic nature in fermions is confirmed with negative spin correlations at nearest neighboring sites and oscillations about $0$ as $n$ increases. The strength of the nearest neighbor spin correlation increases with occupancy. For bosons, the correlations confirm the ferromagnetic nature at $U>V$ and anti-ferromagnetic nature at $U<V$. The spin-chain is able to effectively reproduce the spin and charge correlations obtained from the Hubbard Hamiltonian, producing almost identical plots.

\onecolumngrid

\twocolumngrid

\bibliography{references}

\end{document}